\documentclass{article}
\usepackage[utf8]{inputenc}
\usepackage{fullpage}
\usepackage{appendix}
\usepackage{graphicx}
\usepackage{amsmath}
\usepackage{amssymb}
\usepackage{amsthm, amsfonts}
\usepackage{algorithm, algorithmic}
\usepackage{subcaption}
\usepackage{hyperref}
\usepackage{float}
\usepackage{tikz}
\usetikzlibrary{matrix,positioning}
\tikzset{bullet/.style={circle,fill,inner sep=2pt}}

\newcommand{\R}{\mathbb{R}}
\newcommand{\Z}{\mathbb{Z}}
\newcommand{\N}{\mathbb{N}}

\newcommand{\Expect}{\mathsf{E}}
\newcommand{\Var}{\mathsf{Var}}
\newcommand{\Prob}{\mathsf{Pr}}

\newcommand{\stakedist}{\pi_{\text{stake}}}
\newcommand{\lenddist}{\pi_{\text{lend}}}

\newcommand{\epochstake}{\tilde{\pi}_{\text{stake}}}

\newcommand{\stakeprob}{\hat{\pi}_{\text{stake}}}

\newcommand{\dstake}{\Delta^{\text{stake}}_{i}}
\newcommand{\dlend}{\Delta_{\text{lend}}}

\newcommand{\beginloan}{h_{\text{issued}}}
\newcommand{\termloan}{h_{\text{closed}}}

\newcommand{\wbar}{\overline{\mathbf{w}}}
\newcommand{\mubar}{\overline{\mathbf{\mu}}}

\newcommand{\bmu}{\boldsymbol \mu}

\newcommand{\boSigma}{\boldsymbol \Sigma}
\newcommand{\bphi}{\boldsymbol \varphi}
\newcommand{\bstakeprob}{\boldsymbol \stakeprob}
\newcommand{\bstakeprobi}{\boldsymbol{\hat{\pi}_{\text{stake,i}}}}
\newcommand{\replacement}{\mathsf{R}}

\newcommand{\lstake}{\lambda_{\text{stake}}}
\newcommand{\lslash}{\lambda_{\text{slash}}}
\newcommand{\lborrow}{\lambda_{\text{borrow}}}
\newcommand{\lcollateral}{\lambda_{\text{collateral}}}

\newtheorem{claim}{Claim}
\newtheorem{conjecture}{Conjecture}

\newtheorem{assumption}{Assumption}
\newtheorem{property}{Property}

\usepackage[numbib]{tocbibind}
\usepackage{authblk}
\title{Why Stake When You Can Borrow?}
\author[1]{Tarun Chitra}
\author[2]{Alex Evans}
\affil[1]{{Gauntlet Networks, Inc.
\texttt{tarun@gauntlet.network}}}
\affil[2]{{Placeholder \texttt{alex@placeholder.vc}}}
\date{}
\begin{document}

\maketitle

\begin{abstract}
    As smart contract platforms autonomously manage billions of dollars of capital, quantifying the portfolio risk that investors engender in these systems is increasingly important.
    Recent work illustrates that Proof of Stake (PoS) is vulnerable to financial attacks arising from on-chain lending and has worse capital efficiency than Proof of Work (PoW)~\cite{fanti_pos_econ}.
    Numerous methods for improving capital efficiency have been proposed that allow stakers to create fungible derivative claims on their staked assets.
    In this paper, we construct a unifying model for studying the security risks of these proposals.
    This model combines birth-death P\'olya processes and risk models adapted from the credit derivatives literature to assess token inequality and return profiles.
    We find that there is a sharp transition between `safe' and `unsafe' derivative usage.
    Surprisingly, we find that contrary to~\cite{fanti2019compounding} there exist conditions where derivatives can \emph{reduce} concentration of wealth in these networks.
    This model also applies to Decentralized Finance (DeFi) protocols where staked assets are used as insurance.
    Our theoretical results are validated using agent-based simulation.
\end{abstract}

\section*{Introduction}
Proof of Stake (PoS) is a Sybil resistance mechanism that aims to replace the scarce physical resource usage of Proof of Work (PoW) by using consensus-enforced scarcity of a digital asset.
Moving from PoW to PoS has promised to reduce mining's energy usage, increase scalability, and improve network participation.
PoS achieves this by minting cryptographically-secured tokens according to a fixed monetary policy.
Token holders receive a pro-rata portion of inflation by \emph{staking}, or locking up their tokens in a smart contract, which lets them validate transactions that the network processes.
If the entire token supply is staked, the cost of performing a double-spend attack becomes proportional to 33\% (Byzantine Fault Tolerant, BFT) or 51\% (longest-chain) of the token supply.
Therefore, if the value of a PoS token (relative to a num\'eraire) is large and a large fraction of the outstanding tokens are staked, the network is safe against adversaries with capital proportional to the token's market capitalization~\cite{fanti_pos_econ}.

PoS's threat model is, however, more complex than that of PoW as the usage of capital as a scarce resource leads to novel financial attacks that do not exist in PoW~\cite{fanti2019compounding, chitra2019competitive}.
First, the compounding effects of PoS assets, which do not exist in PoW, make it difficult to design incentive compatible monetary policies~\cite{fanti2019compounding}. 
Moreover, as PoS assets require both market capitalization and staking participation to be high for security, scenarios that keep the market capitalization high, while staking participation remains low can be dangerous to these networks.
The attacks of~\cite{chitra2019competitive} arise in such a scenario, which occur when the system's users and validators are rational profit seeking agents.
One enters this scenario when alternative yields (e.g. lending) exceed PoS returns.
In such scenarios, rational actors unstake their assets and migrate them to alternative vehicles to maximize individual profit, while reducing network security.
This behavior also leads to capital flight and can cause deflationary spirals~\cite{klages2019stability}.

The time value of capital that is lost from locking up an inflationary monetary instrument in a smart contract can be significant, which disincentivizes staking when there exist alternative yield-generating opportunities.
In order to incentivize validator participation, staking protocols have proposed staking derivatives, which allow validators to borrow against their staked assets~\cite{agarwal_ojha_2019, santoni_2019, park, sesameseed_2020}.
This borrowing, which resembles secured lending from fiat finance such as home equity loans, provides a mechanism for validators to gain partial liquidity on their staked capital.
By having the protocol provide lending services (in the form of staking derivatives), one can potentially mitigate the capital flight issues of~\cite{chitra2019competitive}.
For instance, a validator with 1000 tokens of a digital asset $X$ locked in a staking contract can use a staking derivative to mint 750 tokens of a synthetic asset $Y$, representing a borrow of 75\% of stake.
At the beginning of the lien, the synthetic asset could be redeemed one-to-one for the underlying asset.
However, if the validator is slashed, the synthetic asset will redeem for less.
In this case, if the validator is slashed and loses 50 tokens, then they may only be able to redeem 1 $Y$ for 0.5 $X$.

In practice, validators share default risk with other borrowers, which is typically the case when a PoS protocol issues fungible derivative assets, i.e. when all borrowing obligations are denominated in a common asset. 
Such constructions involve issuing staking derivatives backed by the collective obligations of the validator base. Similar aggregations are also common to DeFi lending protocols. In the MakerDAO \cite{mcd_dai} protocol, loans are denominated in a single fungible asset, Dai, whose value depends on a pool of non-fungible loans called ``Vaults." Similarly, lenders in Compound \cite{team2017dai} are issued tokens (called ``cTokens") that entitle holders to a proportional share of the interest generated by all borrowers (for example, lenders of Dai would be issued ``cDai," whose redemption price depends on the interest accrued from all Dai-denominated loans originated by Compound). In both PoS and DeFi, we argue that aggregating obligations from a heterogeneous pool of borrowers into a single asset can be analogized to securitization~\cite{2012securitization}.
The analogy is more than skin-deep, as shown in \S\ref{sec:staking_derivs_main} where we consider the options available to PoS protocols for creating fungible assets representing pools of validator liens.

Unsurprisingly, as the security of a PoS system depends on the total amount of staked assets and their relative value to a num\'eraire (such as US dollars or Bitcoin), this type of borrowing will decrease the security of the PoS system.
Moreover, as illustrated in~\cite{fanti2019compounding}, PoS protocols with improperly chosen monetary policies can lead to concentration of stake in the hands of a few, further reducing network security. 
But how do we quantify this decrease in security?
The above description has numerous parameters, such as how overcollateralized a validator's loan needs to be, what is the interest rate to charge, and how is the redemption pricing curve constructed.
To fully evaluate the safety of the system, one needs to model how the underlying PoS asset, the synthetic derivative asset, and other lending opportunities interact with one another.
In this paper, we will construct a risk model for a broad class of staking derivatives. This model will be analyzed through two lenses:

\begin{enumerate}
    \item \emph{Inequality}: We extend the specialized P\'olya urn model of~\cite{fanti2019compounding} to a generalized process that represents derivatives. Recent advances in probability allow us to \emph{analytically} estimate how derivatives impact inequality.
    \item \emph{Returns and Portfolio Optimization}: Under the assumption of rational validators, we study the returns on portfolios of staked, lent, and derivative assets. We liken staking derivatives to secured loans whose returns depend on staking yields and provide intuition via existing models from quantitative finance.
\end{enumerate}
We will also evaluate these models via numerical simulation and agent-based modeling.

\subsection*{Agent-Based Model}

We extend the agent-based framework of~\cite{chitra2019competitive} to consider rational actors who hold portfolios of staked, lent, and derivative assets.
Participants rebalance their portfolios by maximizing a utility function, which represents the best portfolio that one can hold given current market prices and yields.
Each participant has a different risk preference leading to non-trivial dynamics as riskier borrowers will rebalance their portfolios to be derivative heavy, whereas risk-averse borrowers will have their portfolios be staking coin heavy.
Note that adding a derivative instrument, which represents a leveraged long claim on the underlying PoS token, adds novel risks for validators.
In particular, validators begin to default on these liens when they are slashed for activities that are antithetical to consensus.
Constructing such an instrument involves having the consensus algorithm be aware of the current value of a validator's debt when adjusting the monetary policy of the network.
This is different than the situation of \cite{chitra2019competitive}, where the lending rates are independent of the consensus algorithm used for the PoS asset.

The main machinery that we add to allow consensus to keep track of debt prices and positions is a constant function market maker (CFMM)~\cite{angeris2020improved, angeris2019analysis}.
CFMMs are smart contracts that act as an exchange for the staked coin and the derivative asset.
The usage of CFMMs in cryptocurrencies began with Uniswap~\cite{angeris2019analysis}, which was used for token-to-token exchanges mediated by a CFMM.
Uniswap has had close to \$75 million of digital assets locked in it, demonstrating practical CFMM viability.
However, CFMMs have also been used for other financial applications such as portfolio rebalancing \cite{balancer_wp}, margin trading \cite{futureswap_2020}, and stablecoins \cite{clabscelo, kamvarcelo}.
These simple and versatile mechanisms are parametrized by a convex function that maps asset quantities to an implied price.
This also allows them to be utilized in consensus sensitive matters.
Celo~\cite{kamvarcelo} incorporates a Uniswap-style CFMM in consensus to adjust the PoS protocol's monetary policy based on transaction demand and money velocity.
When the main venue for trading a PoS asset against a lien is an on-chain CFMM, a PoS protocol can adjust its monetary policy and execute margin calls on overleveraged validators.
Ideally, the lending activity related to a PoS token mainly comprises of validators borrowing against their staked assets, allowing consensus to intervene and avoid the scenarios posed in~\cite{chitra2019competitive}. 

If there are a number of defaults in the synthetic asset --- validators borrow against their staked quantity, but cannot repay their loans --- then synthetic asset holders often share the default risk pro-rata.
As a simple example, suppose that we have 10 validators with an equal staked quantity and all of them have borrowed against their stake.
If 20\% of validators default on their loans, then the remaining 80\% have their borrowed assets get 25\% more expensive to close in order to cover the losses of the defaulted loan.
This means that if at time $t_0$, the synthetic asset and the PoS asset have the same price in a CFMM, then at time $t_{\text{default}}$, when both defaults happen, the prices of the synthetic asset should be 75\% that of the PoS asset.
Thus, the exchange rate between the underlying coin and the derivative asset should represent the expected future defaults that the derivative will have to absorb.

We model on-chain lending in the same manner as~\cite{chitra2019competitive} by using the Compound protocol \cite{kao2020compound}.
On-chain lending pools, where lenders share default risk pro-rata, are common in cryptocurrency financial products such as Compound~\cite{kao2020compound}, Uniswap~\cite{angeris2019analysis}, and in PoS itself~\cite{chitra2019competitive}.
Such protocols rely on cryptographic properties of smart contracts to ensure that participants are incentivized to avoid malicious behavior.
This model uses a scoring rule to price the interest rate for borrowing one cryptoasset in exchange for another as collateral based on supply and demand.
If $S_t$ and $D_t$ are the supply and demand (in tokens) for borrowing a token at time $t$, these mechanisms furnish a function $f : \R_+ \times \R_+ \rightarrow [0,1]$ such that $f(S_t, D_t)$ is the interest rate changed to borrowers.
The loans are overcollateralized, like home-equity loans, and the demand is driven by crypto-asset holders who want to borrow fiat currency against their cryptocurrencies for liquidity.
The largest on-chain lending pools on Ethereum are MakerDAO \cite{klages2020stability} and Compound \cite{kao2020compound}, growing to hold hundreds of millions of dollars by early 2020.
These pools have proven to be resilient and provide arbitrageurs with ample opportunity to provide price discovery between lent assets as well as for swaps between pairs of assets.

By combining CFMMs with on-chain lending pools, we are able to focus on how rational agents rebalance their portfolios based on information that is \emph{endogenous to the blockchain itself}.
Our usage of CFMMs implies that as long as one rational arbitrageur exists, on-chain prices will match external prices \cite{angeris2020improved}.
Moreover, the usage of on-chain lending pools implies that all participants can see the same rates for borrowing versus those of staking, allowing for a direct optimization problem to be constructed.
As CFMMs are parametrized by their scoring function $\phi$, it is instructive to consider whether such systems exist in practice.

\subsection*{Staking Derivatives outside of PoS}\label{sec:nonpos}
Existing research on staking primarily focuses on its use as a Sybil resistance mechanism in consensus.
However, Decentralized Finance (DeFi) applications have found novel ways to use staking outside of PoS.
DeFi applications utilize staked capital as `insurance funds' and `security pools' for algorithmic stablecoins, on-chain lending, and margin trading.

Algorithmic Stablecoins are tokens with dynamic monetary policies that adjust to maintain a peg to another asset, such as the US dollar.
These assets increase their token issuance when the synthetic asset is above the peg and decrease it by buying back the stable asset and destroying it (`burning').
Protocols such as Celo~\cite{clabscelo} and Terra~\cite{terra_wp} utilize a staking token that represents both on-chain transaction fees as well as a debt instrument.
In these protocols, stakers earn interest for transaction validation, but can have their assets diluted in order to buy back the stable asset when it is trading below its peg.
Celo~\cite{clabscelo} and Terra~\cite{dimaggio_2019} use CFMMs to handle this dilution process. 
This usage of CFMMs to enforce protocol constraints fits into the framework presented in \S\ref{sec:general_model}.

On-chain lending protocols, such as Aave~\cite{aave_risk} and Compound~\cite{compound2019, kao2020compound}, have insurance funds that take a fraction of earned interest and hold it in escrow.
This earned interest is used to pay back lenders when external market conditions cause failure modes such as cascading liquidations~\cite{kao2020compound}.
These failure modes represent edge cases where a decentralized lending protocol is unable to sell assets to cover liabilities due to market volatility.
MakerDAO's `Black Thursday' events on March 12, 2020 represent a realized instance of this situation~\cite{whiterabbit_2020}.
As lending protocols aim to decentralize their ownership, new token models from Aave~\cite{turley_2020} and Compound~\cite{leshner_2020} require staking for participation in both the governance and insurance functions of the system.
In this use case, the insurance fund pays out stakers from received interest and dilutes the staking token when mass defaults happen.
The lending contract effectively securitizes the insurance fund via this mechanism, allowing for a larger pool of capital to back the algorithmic loans issued.
These mechanisms resemble reinsurance and can also be analyzed using the staking derivatives framework.

Margin trading protocols such as Synthetix \cite{synthetix} and UMA \cite{uma} utilize a staking token that acts akin to an exchange seat at the CME.
When this token is staked in Synthetix, it is eligible for both staking rewards and can be used as collateral for a synthetic asset.
If a user wants to mint a synthetic version of the S\&P 500, they need to provide a price feed and collateral (in the staking asset).
Once minted, users can place pairwise margin trading bets on whether the S\&P 500 will go up or down, with the shorts paying the longs if the asset moves up.
These synthetic assets have risks that arise when the collateral value falls precipitously, leaving the system unable to pay the winning side of a margin trade.
Thus, similar to the on-chain lending case, the staking token allows investors to earn interest when trades are settled, but possibly absorb losses when the system is unable to pay out as expected. This is another form of a staking derivative that falls within the framework of \S\ref{sec:general_model}.

Finally, we note that Tezos's `virtual baker' mechanism \cite{gold_2020} as well as the stochastic model of MakerDAO \cite{klages2020stability} both fit into our framework, with different choices of CFMM.
We will map these various uses of staking derivatives to a CFMM pricing function in \S\ref{sec:deriv_ex}.
The variety of uses of staking derivatives in PoS and DeFi suggest that a pricing model that covers known use cases and is simple to interpret will be valuable for the evolution of these systems.
The remainder of this paper will describe the assumptions and model used to characterize staking derivatives.
We will prove some properties about this model and use numerical simulation to check that these transitions exist.
Finally, we conclude with a discussion about how to instantiate the different models discussed in this section via choices of the CFMM pricing function $\phi$.

\subsection*{Takeaways for PoS \& DeFi protocol designers}
We highlight the main results that are relevant for DeFi and PoS protocol designers.
First, the results of \S\ref{sec:stake_derive} show that one can reduce inequality amongst stakers by adding staking derivatives.
However, the portfolio returns estimated in \S\ref{sec:three_comp_results} show that this inequality comes at the cost of burning a significant fraction of the money supply.
These two results describe a trade-off between the number of liquidations of a derivative position and the amount of inequality in the system.
The phase transitions estimated via simulation in these sections suggest that there is a small region of parameter space where one can balance these trade-offs optimally.
In this `optimal region,' one can use a staking derivative to be the dominant form of borrowing in the system, so that the on-chain lending attacks of \cite{chitra2019competitive} are ineffective.
This optimal region is the portion of the `safe' regime that is closest to the phase transition boundary (see Figures \ref{fig:Gini_mean}, \ref{fig:stake_weight} for examples).
Thus for staking systems that admit significant leverage (e.g. Synthetix \cite{snx_litepaper}), it can be crucial to ensure that the system stays within this region.
While the analytical results of \S\ref{sec:stake_derive} and \S\ref{sec:three_comp_results} provide bounds on these transitions, the precise boundary between the `safe' and `unsafe' region which is optimal can only be explored via simulation.
Figures \ref{fig:final_borrow_weight}, \ref{fig:supply_ratio_const} illustrates how the size of a validator's slashed bond affects the size of the optimal region.

The risk models used in \S\ref{sec:stake_lend_deriv} for the `safe' derivative region are analogues of models used to analyze credit derivatives, such as mortgage-backed securities (MBS).
In the MBS literature, one considers derivative securities with `embedded options.'
These options correspond to homeowners deciding to prepay their mortgage (leading to a loss of interest) or refinancing their mortgage.
Accurately pricing MBSs relies on including the value at risk to an investor when these options are exercised.
Staking derivative loans can also be thought to have embedded options that are granted to the lender (the PoS protocol) instead of the borrower.
These may involve default protections such as penalizing validators whose derivative is deemed overleveraged, or otherwise adjusting the derivative pricing function $\phi$ on a per-user basis.
Designers can embed a notion of a `credit score' in $\phi$ that allows for borrowers with good borrowing histories to mint derivatives with lower fees.
Given that a number of DeFi protocols have proposed including similar benefits to active participants \cite{aave_lps, snx_liqrewards}, it is important to correctly value these embedded options when choosing pricing curves.
Claim \ref{claim:lending_supply} provides a technique for a wide class of staking derivatives (including all proposals in DeFi and PoS known to the authors) for approximating the risks associates with such options.
The results of \S\ref{sec:sld} extend credit derivative risk measures \cite[Ch. 6]{davidson2014mortgage} to staking derivatives and can be used to choose $\phi$.
\subsection*{Outline}
The remainder of the paper will focus on describing the staking derivatives framework.
We note that mathematical notation and details about the assumptions made about the Proof of Stake model, based on those from \cite{chitra2019competitive}, can be found in Appendices \ref{sec:notation} and \ref{sec:assumptions}.
The main new mathematical object introduced in this paper, the \emph{derivative pricing function} $\phi$ for CFMMs, will be introduced in \S\ref{sec:general_model}.
We construct derivative pricing functions for both PoS networks and for DeFi protocols that utilize staking mechanisms.
The main difference is that DeFi protocols rely on different boundary conditions than those of PoS networks. 
Subsequently, we will construct two models that utilize $\phi$ to price derivatives: an urn model and a portfolio risk model.

The urn model of \S\ref{sec:stake_derive} will provide a way to measure concentration of wealth and inequality in the stake distribution.
In this section, we will use theoretical results on continuous-time embeddings of urn models to measure how derivatives affect concentration of wealth.
Surprisingly, our main result is that there exist scenarios where the existence of derivatives lowers the amount of inequality in the system.
We illustrate this result theoretically by using a proxy for the Gini coefficient, the ratio of $L^2$ to $L^1$ norms of the stake distribution.
Subsequently, we use a simulated agent-based model to numerically verify this result under more realistic conditions and verify that the Gini coefficient indicates a less concentrated stake distribution. 

Portfolio risk is assessed in \S\ref{sec:portfolio_sel}.
The main results that we find are that there are `safe' and `unsafe' regimes for both of these systems, which are characterized by features of $\phi$ and the monetary policy of the network.
When in the safe regime, we show that this model can be interpreted in a manner similar to credit derivatives with embedded options.
This interpretation allows for us to compute the expected returns for both the validator and the protocol, as they hold portfolios of staked and borrowed assets.
We further show that in the safe region, the results of \cite{chitra2019competitive} on capital flight due to lending still hold.

\section{Derivative Pricing Function}\label{sec:general_model}
The main object of study in this paper is the derivative pricing function for validator $i$, denoted $\varphi_i$.
This function allows for pricing synthetically minted assets in terms of an underlying asset.
In order to understand how to construct $\varphi_i$, we will first need to study the function of staking derivatives.

\begin{figure}
    \centering
    \includegraphics[scale=0.6]{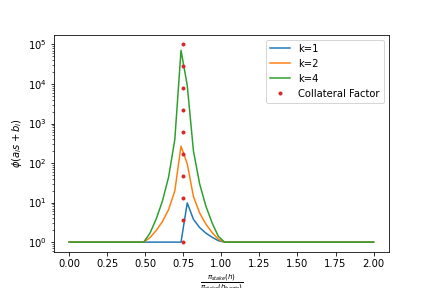}
    \caption{$\varphi_i$ constructed from $\phi(s) = \frac{1}{s^k} \wedge 1$ for different exponents $k$. The $x$-axis is the ratio $\frac{\stakedist(h)}{\stakedist(\beginloan)}$. This ratio only decreases when a validator is slashed in PoS or if a price moves against a direction in DeFi (e.g. sUSD/SNX price goes down, even though a staker is long). The dotted line represents a collateral factor of $c_i = 0.75$. The loan is liquidated, represented by $\phi(c_i\stakedist(\beginloan)) = \infty$. The steepness in the changes in $\phi$ from the different exponents control how leveraged a validator is as they are increasingly slashed.}
    \label{fig:phi}
\end{figure}

\subsection{Staking Derivatives}\label{sec:staking_derivs_main}
Consider the Tezos PoS network \cite{goodman2014tezos}, which has been live since 2018 and whose XTZ token is the largest PoS coin by market capitalization.
Suppose that Tezos validators were allowed to borrow up to 75\% of their staked assets in the form of a synthetic XTZ, sXTZ, with value equal to the market price of XTZ at inception.
In order for a validator to recover their stake and earned block rewards, they need to buy back their XTZ with sXTZ.
Such a synthetic asset is known as a \emph{staking derivative}, as it represents a lien against staked assets \cite{agarwal_ojha_2019}.
Unlike other on-chain liens,\footnote{In MakerDAO, Compound, and Synthetix, the protocol pays liquidators to buy defaulted liens, whereas a PoS currency can simply burn a validator's stake} this liability is known to the consensus protocol, which can ensure that if a validator has less stake than what they borrowed, they lose their assets.
Note that validators who use such liens to gain liquidity are taking leveraged long positions on the underlying asset (e.g. XTZ), akin to a ``Vault" in MakerDAO \cite{klages2020stability}.

For example, suppose that a validator has 10,000 XTZ staked in the network with a collateral factor of 75\%, meaning they can borrow up to 7,500 XTZ worth of sXTZ.
If the current price of sXTZ is 1.01 XTZ, then a validator borrowing maximally against 1,000 XTZ would receive $1,000 \times 1.01 \times 0.75 = 757.5$ sXTZ.
The borrowing validator can use the sXTZ as a normal token and sell it (e.g. for a stablecoin or Bitcoin), allowing them to have increased capital efficiency and liquidity.
However, in order for the borrowing validator to reclaim their staked XTZ, they have to repurchase their stake using sXTZ.
If the validator borrowed 1,010 sXTZ but was slashed, the price in sXTZ that they have to pay the network to reclaim his collateral is higher than 1,010 sXTZ.
Moreover, if the validator is slashed by 2,500 sXTZ (e.g. they have less staked than their 7,500 limit), then their loan is defaulted on and the network reclaims 1,010 sXTZ.
On the other hand, if the validator's stake upon repaying the borrowed asset is greater than or equal to the quantity staked when the loan was taken out, the borrower's price remains 1,010 sXTZ.
These constraints ensure that the PoS network penalizes validators who borrow against their staked assets and perform a malicious activity (that leads to slashing).

Let $\beginloan$ be the block height that validator $i$ minted a staking derivative against their stake at that time $\stakedist(\beginloan)_i$ and let $\termloan$ be the block height they closed their loan.
Moreover, let $c_i \in [0, 1]$ be the $i$th validator's \emph{collateral factor}.
This represents the fraction of stake that can be borrowed against, e.g. $c_i \stakedist(h)$ is the fraction allowed to be borrowed.
For a staking derivative to be economically sound or \emph{solvent}, the following properties are necessary:
\begin{property}\label{prop:default}
    \textbf{Default if overleveraged}. If a validator borrows at block height $\beginloan$, but at height $h' > \beginloan, \stakedist(h')_i < c_i \stakedist(\beginloan)_i$, then the network can reclaim the validator's stake and redistribute it as fit.
\end{property} 
\begin{property}\label{prop:repay}
    \textbf{Repayment amount bounded below}. If a validator mints $x$ sXTZ at block height $\beginloan$ then no matter what, they will always at least pay $x$ sXTZ to regain their collateral. This ensures that the sXTZ/XTZ price will be greater than or equal to 1, meaning that the synthetic will never be worth more than the underlying asset.
\end{property}
\begin{property}\label{prop:mono}
    \textbf{Monotonically increasing payment}. Suppose a validator borrows $x$ sXTZ at block height $\beginloan$. Suppose they have to pay back $x(i)$ sXTZ if they are slashed $i$ times. The borrowing mechanism is said to be \emph{monotonic in payment} if $x(i)$ is increasing in $i$ --- the more you are slashed, the more you have to repay.
\end{property} 
\begin{property}\label{prop:onch}
    \textbf{There exists an on-chain synthetic-to-real market}. In the previous example, we require the sXTZ to XTZ price to mark the loan and to figure out much a validator needs to repay. In order for the consensus algorithm to execute a default, it needs this market price. Moreover, borrowers need access to this market to purchase sXTZ to close their liens.
\end{property}
These conditions ensure that the protocol mints derivatives that are always solvent.
Note that these conditions are analogous to those required by over-collateralized lending contracts \cite{kao2020compound} and by debt-driven algorithmic stablecoins \cite{klages2019stability, klages2020stability}.
The first condition ensures that the system's assets (e.g. the staked coins) are always greater than net liabilities (e.g. the derivative).
The second condition ensures that there is no economic abstraction of the underlying asset by the synthetic asset. 
In particular, this implies that market will never find sXTZ to be more valuable than XTZ, which could lead stakers to borrow sXTZ against their staked XTZ and then default.
Monotonically increasing payments ensure that bad borrowers (e.g. validators who are repeatedly slashed) have to pay the network more to borrow against their stake, as they are providing riskier collateral to the network.
Finally, the last condition ensures that borrowers can easily buy sXTZ to repay their debt and ensures that the protocol can correctly adjust the monetary supply upon realizing a default. 

Note that while it is possible to replace the final condition with the existence of a decentralized oracle \cite{peterson2015augur} instead of an on-chain market, we argue that this is not feasible for staking derivatives.
Firstly note that decentralized oracles such as Augur \cite{peterson2015augur} or Chainlink/DECO \cite{zhang2019deco} rely on smart contract protocols for their execution.
Since staking derivatives are used for the base protocol's staking asset, the smart contract would need to run on the same network, which would cause a variety of issues.
For instance, the oracle contract could be censored by validators who have large liens that are in default, which is a form of 'miner extractable value' \cite{daian2019flash}. 
Moreover, if an oracle were used, the PoS consensus protocol would be subject to manipulation from this oracle, as it will have security under a different threat model.
Finally, we note that while CFMMs can be manipulated, it is expensive to do so.
The cost of manipulating a CFMM is linear in the size of the liquidity pool, whereas oracle manipulation has a constant cost \cite[Appendix E]{angeris2019analysis}.

How can we enforce these constraints within the protocol?
Constant function market makers (CFMM) \cite{angeris2019analysis, angeris2020improved} provide on-chain mechanisms for pricing baskets of assets based on quantities deposited by participants.
These mechanisms rely on two principal agents: liquidity providers (LPs) and traders.
Liquidity providers lend their assets to a smart contract and upon each trade executed by a trader, receive a pro-rata share of transaction fees.
The main parameter needed to configure a CFMM is a scoring rule $\Phi : \R_+^k \rightarrow \R_+$ that maps a vector of quantities of assets $q \in \R_+^k$ to an invariant.
Provided that $\Phi$ is closed and convex, one can compute the prices of any pairs of assets. For staking derivatives, we require a function $\Phi : \R_+^2 \rightarrow \R$ that can provide the price of the synthetic asset relative to the underlying asset (e.g. sXTZ/XTZ) and satisfy the above constraints.
We note that when constructing two-asset CFMMs, it is sufficient to provide a pricing curve $\varphi: \R_+ \rightarrow \R_+$ to construct a CFMM $\Phi$ \cite{angeris2019analysis}. 
We provide an explicit example of such a construction in the sequel. 

\section{Constructing $\varphi$}\label{sec:construct_phi}

Before constructing $\varphi$, let us first look at how it is used to close out liens.
Suppose that we have a validator who has staked assets $\stakedist(h)_i$ at block height $h$, then the validator is allowed to use $\varphi$ to mint up to $c_i \stakedist(h)_i$ in synthetic assets.
Therefore, each validator can be thought of as holding a portfolio $\Pi(i, h) = (\stakedist(h)_i, -\delta_i)$ of staked assets at block height $h$ $\stakedist(h)_i$ and synthetic borrowed assets $\delta_i$.
In order to close out the $-\delta_i(h)$ position, the validator has to pay the borrowing contract $\varphi(\stakedist(h)_i)$.
The first three conditions from the previous section are represented as follows:
\begin{itemize}
    \item \textbf{Default if overleveraged}: If $\varphi(\stakedist(h))_i < c_i \varphi(\stakedist(\beginloan))_i$ then $\Pi(i, h) = (0, 0)$
    \item \textbf{Repayment bounded below}: For all $h > \beginloan\;\varphi(\stakedist(h)_i) \geq \delta_i$
    \item \textbf{Monotonically increasing payment}: If for $h, h' > \beginloan$ we have $\stakedist(h)_i < \stakedist(h')_i$ then $\varphi(\stakedist(h)_i) \geq \varphi(\stakedist(h')_i)$
\end{itemize}

Recall that in Assumptions \ref{ass:slashing_prob} and \ref{ass:collat}, we assume that each validator has their own collateral factor $c_i$ and probability of being slashed $p_i$.
This means that each validator should have a pricing function $\varphi_i$ that depends on their collateral factor and likelihood of being slashed.
To make the borrowed asset fungible (e.g. similar to Compound's wrapped tokens, such as \verb=cDAI=), we will need to construct $\varphi$ as an aggregation of individual validator's pricing functions $\varphi_i$.
Note that this is analogous to the `non-fungible' CDPs in MakerDAO which give rise to a `fungible' synthetic token Dai.
The precise form of the aggregation, such as the bounded mean aggregation $\varphi(\stakedist(t)) = \frac{1}{n} \sum_i (\varphi_i(\stakedist(t)_i) \wedge \varphi_{\max})$ or $\varphi(\stakedist(t) = \mathsf{Median}(\varphi_1(\stakedist(t)_1), \ldots, \varphi_n(\stakedist(t)_n))$, does not affect our
results\footnote{The form of the aggregation chosen will affect liquidity and fees, but as long as the aggregation satisfies the necessary and sufficient conditions of \cite{angeris2020improved}, we can still use $\varphi$ as a CFMM} and for the remainder of the paper we will focus on dealing with $\varphi_i$. 

We can think of the synthetic asset as ``shares" collateralized by the assets staked in the network.
These shares, which are tradeable against the underlying staked asset provide the validator with liquidity that depends on their collateralization ratio $c_i$ and their current stake.
Thus, $\varphi_i$ takes in a validator's current stake and the stake they had when they borrowed and returns the number of shares need to buy a single staking token.
Intuitively, when a validator has more stake in the network (inclusive of rewards earned since the loan was issued) than when she initially minted shares, the validator should be able to redeem one share for one staking token. 
Moreover, when a validator has defaulted --- their current stake is less than $c\%$ of the stake when the shares were issues --- the price should be infinite.
Thus, if we let $\beginloan$ be the block height at which a loan was issued, we have the constraints 
\begin{align}\label{eq:bdy_conditions}
    \varphi_i(\stakedist(h)_i) &= 1 & \text{if }\stakedist(h)_i > \stakedist(\beginloan)_i \nonumber \\
    \varphi_i(\stakedist(h)_i) &= \infty &  \text{if } \stakedist(h)_i < c_i \stakedist(\beginloan)_i  
\end{align}
The specified boundary conditions encapsulate properties \ref{prop:default} and \ref{prop:repay}.

In order to fully specify the model, we have to describe $\varphi_i$ that satisfy the boundary conditions \eqref{eq:bdy_conditions} and property \ref{prop:mono}.
We will assume that for all validators $i, \varphi_i(s) = \phi(a_i s + b_i)$ where $\phi$ is a `mother' valuation function\footnote{This terminology is adapted from the wavelet literature, where there is a `mother' wavelet function that determines all of the spatiotemporally localized basis functions \cite{daubechies1992ten}} and $a_i, b_i$ are coefficients that are chosen to ensure that we satisfy \eqref{eq:bdy_conditions}.
Furthermore, we will assume the following properties of $\phi : \R_+ \rightarrow \R_+$:
\begin{enumerate}
    \item $\phi$ is continuous on its domain, e.g. $\phi \in C^0(\R_+)$
    \item $\phi$ restricted to $(0,1)$ is smooth, e.g. $\phi\vert_{(0,1)} \in C^{\infty}((0, 1))$
    \item $\phi$ is decreasing in its argument, $\forall x \in [0, \infty),\, \phi'(x) \leq 0$
    \item Boundary conditions: $\phi(0) = \infty, \phi(1) = 1$
\end{enumerate}
Note that the second property, $\phi'(x) \leq 0$ forces us to satisfy property \ref{prop:mono}, as desired.
A simple example of an admissible $\phi$ is $\phi(s) = \max(s^{-1}, 1) = s^{-1} \vee 1$.
The former assumption ensures that outside of the boundary conditions, we have a relatively easy to deal with valuation function whose expectation can be computed easily.
Other the other hand, the latter encodes the idea that as a validator has more stake in the network, relative to their borrowed quantity, they should have to pay a decreasing number of shares to recover their collateral.
Furthermore, we note that the boundary conditions \eqref{eq:bdy_conditions} are satisfied if we set $a_i$ and $b_i$ as follows:
\begin{align}\label{eq:affine}
    a_i &= \frac{1}{\stakedist(\beginloan)_i (c_i-1)} \\
    b_i &= \frac{c_i}{1-c_i}
\end{align}
Note that with these parameters, $a_i s + b_i = 0$ when $s = c_i \stakedist(\beginloan)_i$ and $a_i s + b_i = 1$ when $s = \stakedist(\beginloan)_i$.
Figure \ref{fig:phi} provides an example of this for $\phi(s) = \frac{1}{s^k} \wedge 1$, where you can see how the coefficients $a_i, b_i$ shift the the curve $\phi$ to the curve $\varphi_i$, with $c_i = 0.75$.

Such linear aggregations describe a number of real-wold DeFi protocols such MakerDAO, wherein a heterogeneous pool of loans based on different collateral assets is aggregated to a fungible asset, Dai~\cite{mcd_dai}. The obligations of all borrowers are thus denominated in the same asset. The aggregation can be thought of as securitization~\cite{2012securitization,securitization_eur} where $\phi$ provides exposure to a pool of stake-collateralized loans represented by $\phi_i$.

\subsection{Other Examples of Derivative Pricing Functions}\label{sec:deriv_ex}
This form of the pricing function provides a general model of a redemption curve for synthetic liens that can be adapted to support a variety of applications. For completeness, we illustrate with some examples of real-world staking applications, including both PoS protocols and DeFi applications (where staking is not used for consensus). 

In the Synthetix protocol, users stake Synthetix Network Tokens (SNX) which serves as collateral for the issuance of synthetic assets.
The sum of all outstanding synthetic assets represents the debt of the system the risk of which is shared among all SNX staked \cite{snx_litepaper}.
The security of the Synthetix protocol therefore benefits from maintaining both higher value and staking participation in SNX.
In exchange for serving as the counterparties for synthetic asset exchanges, SNX stakers are rewarded with transaction fees generated from the Synthetix exchange as well as new inflation.
In addition, to encourage liquidity for synthetic assets, Synthetix rewards users who contribute synthetic tokens to CFMMs such as Uniswap and Curve \cite{snx_liqrewards}.
The scenario where SNX stakers borrow synthetics against their staked supply and `lend' these to an on-chain CFMM corresponds to the model in \S\ref{sec:stake_lend_deriv}.
In this scenario, $\phi_i$ is simply the ratio of the price of SNX to the price of the synthetic asset times a user's stake.
The liquidity reward is added to the return on lending market, in this case the CFMM.  
From Synthetix's smart contracts \cite{snx_github_repo}, one finds that $\varphi_i(\stakedist(h)_i) = \frac{P_{SNX}}{P_{synth}}$ when the agent is sufficiently collateralized, where $P_{SNX}$ and $P_{synth}$ are the prices of SNX and the synthetic respectively. Mean lending return is $\gamma_t = \gamma_{b_t} + R\frac{\lambda_i}{\lambda} $ where $\gamma_{b_t}$ is the expected return offered by the CFMM, R is the absolute SNX liquidity reward and $\frac{\lambda_i}{\lambda}$ is the relative amount lent by agent $i$ to total lent assets.
Note that with this choice of $\phi$, we need to relax the first boundary condition of equation \eqref{eq:bdy_conditions} to allow $\varphi_i(\stakedist(h)_i) \leq 1$.
This difference in boundary condition does not mutate any of the formal results in \S\ref{sec:stake_derive}, but does affect the results of \S\ref{sec:sld}.
We also note that decentralized derivatives from the Vega protocol \cite{danezis2019vega} have a pricing mechanism similar to $\varphi_i$ that has been studied and found to have stable Nash equilibria \cite{siska2020incentives}.

In the Tezos protocol, stakers (termed ``bakers") may choose to provide their assets to on-chain CFMMs while retaining a portion of their staking rewards \cite{dexter}. In effect, forgone staking returns can be seen as an interest rate for borrowing against staked assets in order to supply them to on-chain lending (the CFMM). For positive reductions in staking reward, $\phi_i$ is a submartingale whose positive drift is determined by foregone staking return. In the Polkadot~\cite{polkadot_wp} PoS system, staking derivatives are managed by an independent protocol that accepts deposits of staked PoS tokens (DOTs) and issues staking derivatives (L-DOTs) in exchange~\cite{LDOTs}. Staking-as-a-service businesses wherein users delegate their stake and outsource staking operations to a third party may also issue staking derivatives to their clients, as discussed in \cite{harmony_seed} in the case of the Harmony PoS protocol. Such examples are simpler to analyze, as default risks are concentrated in the customer base of the service provider.

A construction of an explicit redemption price between synthetic and staked assets for second-layer proof-of-stake protocols on Ethereum is considered in \cite{curve_nucypher}. Such protocols may include L2 sidechains, DeFi protocols, and off-chain privacy protocols that require main-chain staking such as~\cite{keep,nucypher}. In this proposal, an on-chain CFMM is endowed with a portion of the PoS protocol's native assets, which are used to facilitate exchange between the staking asset and a staking derivative. The CFMM automatically enforces a target value for $\phi$, which is typically growing uniformly with time. When the staking derivative asset is undervalued relative to the target price, staking revenues from the CFMM's reserves are diverted to purchasing the synthetic. Similarly, if the synthetic is overvalued relative to the target, reserves are sold to the CFMM to restore the price. Note that in this work we study the resulting price process $\phi$ and our approach is agnostic to the particular enforcement mechanism.

DeFi lending protocols have begun to offer credit using reserves supplied to CFMMs as collateral. For example, Aave allows a user who has supplied reserves to a Uniswap pool to gain partial liquidity on their assets by borrowing using liquidity pool shares as collateral~\cite{aave_lps}. While these examples do not explicitly involve PoS and are thus outside the scope of this work, our results can easily be adapted to similar applications.

\section{Concentration of Wealth}\label{sec:stake_derive}
We will first consider a two-component model where each agent is represented as a validator whose assets are either staked or borrowed.
This model will provide a stochastic process that evolves the stake distribution $\stakedist(h) \in \Delta^n$ and the derivative distribution $\delta(h) \in \Delta^n$.
We extend the simple PoS models of \cite{chitra2019competitive, fanti2019compounding} for each validator $i$ and block height $j$, where a single validator is chosen as a block producer and given a reward.
In our extension, we assume that a validator loses a fraction $\iota \in (0, 1)$ of their stake when slashed.

\begin{figure}
    \centering
    \includegraphics[scale=0.3]{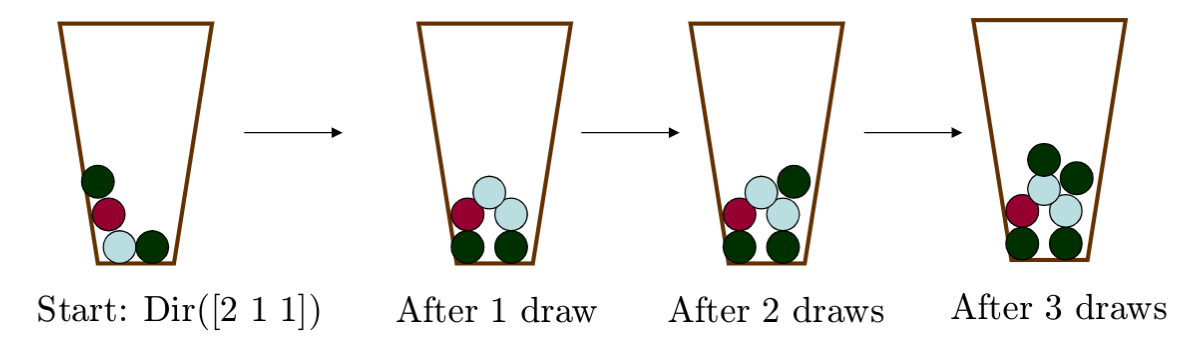}
    \caption{Figure of a P\'olya urn sample process \cite{zandie_2019}. The initial distribution of balls follows a $\mathsf{Dirichlet}(2, 1, 1)$ distribution. After the first ball is drawn (light blue), another one is added in. This process is repeated to get to the final state. Note that the convergence to $\mathsf{Dirichlet}(2, 1, 1)$ comes from \cite[\S3.1.2]{ghosh2003bayesian}}
    \label{fig:polya}
\end{figure}

\subsection{Probability Space}\label{sec:prob_space}
In order to describe the probabilistic nature of slashing and how it affects defaults, we define a probability space via the state transitions that occur when a validator is slashed.
Let $p_i$ be the probability that the $i$th validator is slashed (c.f. Assumption \ref{ass:slashing_prob} in Appendix \ref{sec:assumptions}).
For our scenario there are four outcomes for validator $i$ at block height $h$:
\begin{enumerate}
    \item Rewarded and Not Slashed, $\mathcal{E}_1$: This leads to a $\stakedist(h)_i = \stakedist(h-1)_i + R_h$ and occurs with probability $\Prob_h[\mathcal{E}_1 | i] = (1-p_i) \cdot \stakedist(h)_i$
    \item Not Rewarded and Not Slashed, $\mathcal{E}_2$: This leads to no change in stake, $\stakedist(h) = \stakedist(h-1)$ and occurs with probability $\Prob_h[\mathcal{E}_2 | i] = (1-p_i) (1-\stakedist(t)_i)$
    \item Slashed but no default, $\mathcal{E}_3$: This leads to a change in stake of $\stakedist(h)_i = (1-\iota)\stakedist(h-1)_i$ and occurs with probability $\Prob_h[\mathcal{E}_3 | i] = p_i (1-\mathbf{1}_{\stakedist(h)_i < c_i \stakedist(h)_i})$
    \item Slashed and defaulted, $\mathcal{E}_4$: This leads to a stake change of $\stakedist(h)_i = 0$ which occurs with probability $\Prob_h[\mathcal{E}_4 | i] = p_i\mathbf{1}_{\stakedist(h)_i < c_i \stakedist(h)_i}$
\end{enumerate}
We denote the set of possible outcomes for a validator as $\mathcal{E} = \{\mathcal{E}_1, \mathcal{E}_2, \mathcal{E}_3, \mathcal{E}_4\}$.
As the probabilities of these events change as a function of block height, we have an infinite sequence of probability measures $\Prob_h$ on $\mathcal{E}$. 
Note that if you are slashed, it overrides whether you won a reward or not, matching the policies of the two largest staking networks, Tezos and Cosmos.

Unlike the compounding block rewards of \cite{fanti2019compounding}, there are a number of differences when slashing is introduced.
Without slashing, the stochastic evolution of $\stakedist(h)$ is increasing in that the money supply at height $h, S_h = \Vert \stakedist(h) \Vert_1$, is increasing in block height $h$.
With slashing, this is not true, as the stake lost due to slashing makes the monetary supply non-monotonic.
However, the model contained in \cite{fanti2019compounding} considers a situation where a selfish mining adversary causes the monetary supply to be non-increasing as a function of height.
The authors of \cite{fanti2019compounding} analyze this by considering the evolution of $\stakedist(h)$ as a P\'olya urn process and add in adversarial behavior via what is termed\footnote{This `time-dependence' can be characterized by a P\'olya urn that allows for the removal of balls, as in the model presented in the next section} `time-dependence'.
In order to handle slashing, we generalize this P\'olya urn model to handle the removal of stake in a manner that is a superset of the adversarial scenario of \cite[\S4]{fanti2019compounding}.

\subsection{Urn models}
Urn models were first introduced by P\'olya and Eggenberger in 1923 \cite{polya_original} to study contagions in epidemiology.
These models have been used in a variety of fields, including in the analysis of randomized branching algorithms \cite{dembo2020everything, roesler2001contraction} that are common in blockchains.
The simplest urn model considers an urn filled with $r$ red balls and $g$ green balls.
A ball is drawn from this urn and another ball of the same color is added.
For instance, if a red ball is drawn with probability $\frac{r}{r+g}$, then another red ball is added, so that the probability of a subsequent red draw is $\frac{r+1}{r+1+g}$ (see figure \ref{fig:polya} for a reference).
The sample paths of this process exhibit the `rich-get-richer' phenomenon for certain initial conditions, where the urn ends in a state with one dominant color.
Urn models have been studied with systems that have $n$ balls and such that the sampling process is enriched with more complex replacement strategies.
Moreover, these models serve as the prototype for exchangeable random processes, which are permutation-invariant but not uncorrelated stochastic processes.\footnote{de Finetti's theorem says that all exchangeable stochastic processes are representable via a sequence of urns with a particular replacement strategy \cite{mahmoud2008polya}}

In \cite{fanti2019compounding}, the authors model a PoS system as an urn with balls of $n$ different colors.
Each ball represents a validator and the initial stake distribution $\stakedist(0)$ represents the number of balls of each color at the chain's genesis.
When a ball of color $c$ is selected, $R_h$ $c$-colored balls are added to the urn.
In the adversarial scenario of \cite[\S3]{fanti2019compounding}, a selfish validator represented by color $c$ can cause the monetary supply to remain unchanged\footnote{This is because the selfish mining adversary causes an honest participant's block to become an orphan, leading to a loss of block reward for the honest miner and a gain for the adversary} even though a new block is mined.
The selfish validator coerces the sampling procedure into giving the selfish validator $2R_h$ $c$-colored balls while removing $R_h$ $c'$-colored balls, where $c'$ is the color of an honest validator.
This occurs when a selfish validator publishess two blocks whose root is at height $h_{\text{current}}-1$ which invalidates an honest validator's block at the current block height $h_{\text{current}}$ as the selfish validator's chain is longest.

Generalized urn models allow for balls of color $c$ to affect the concentrations of balls of other colors.
These models are specified by a \emph{replacement matrix} $\replacement \in \Z^{n \times n}$.
The entry $\replacement_{c,c'}$ is the number of balls of color $c'$ to add when ball $c$ is drawn.
When $\replacement_{c, c'} < 0$, we remove $\replacement_{c, c'}$ balls of color $c'$ from the urn when a ball of color $c$ is drawn.
This allows for the urn model to act like a birth-death process, where certain draws of one color reduce the likelihood of other colors being drawn in the future. 
In the previous adversarial validator example, if we only have two validators ($n=2$), then the replacement matrix for the selfish mining strategy is,
\begin{equation}\label{eq:adv_matrix}
\replacement = 
\begin{bmatrix}
    2R_h & -R_h \\ 
    0 & R_h
\end{bmatrix}
 = 
 R_h
 \begin{bmatrix}
    2 & -1 \\ 
    0 & 1
\end{bmatrix}
\end{equation}
where the first row represents an adversary's draw, while the second row represents the honest participant.

Suppose that we initially start with a stake distribution $\stakedist(0) \in \Z^n$.
At block height $h$, a validator $v(h)$ is selected based on the stake distribution, e.g. $v(h) \sim \stakeprob(h)$.
Then, we update the stake distribution as:
\begin{equation}\label{eq:stake_update}
\stakedist(h) = \stakedist(h-1) + \replacement_{v(h)}
\end{equation}
where $\replacement_{v(h)}$ is the $v(h)$-th row of $\replacement$.
With this framework, there are a number of results that provide limit laws for the distribution of terminal stake, $\bstakeprob = \lim_{h\rightarrow\infty} \stakeprob(h)$, which depend on $\replacement$.
For instance, if $\replacement = sI$, e.g. a constant multiple of the identity matrix, then $\frac{\stakedist(h)}{hs} \rightarrow \mathsf{Dirichlet}(\stakedist(0))$, where the convergence is in distribution \cite{mahmoud2008polya, janson2019rate}.
On the other hand, if we perform this update with the replacement matrix of eq. \eqref{eq:adv_matrix}, then most of the results of \cite{fanti2019compounding} (e.g. the stochastic domination results for certain strategies) are direct corollaries of the birth-death limit laws of \cite[Theorems 1, 2]{kuba2011death}.

As described in \S\ref{sec:prob_space}, slashing introduces another set of non-determinism that affects replacement.
To incorporate this effect, we will need the replacement matrix to be \emph{random} and respect the probabilities described in \S\ref{sec:prob_space}.
This means that each row will be drawn from a distribution, or equivalently, each row is a probability measure.
In particular, the $i$th row of a replacement matrix with slashing has the form:
\begin{align}\label{eq:replacement_measure}
\replacement_i &= \Prob_h[\mathcal{E}_1 | i] \delta_{R_h} + \Prob_h[\mathcal{E}_2 | i] \delta_{0} \nonumber \\
&+ \Prob_h[\mathcal{E}_3 | i] \delta_{-\iota\stakedist(h-1)_i} + \Prob_h[\mathcal{E}_4 | i] \delta_{-\stakedist(h-1)_i}
\end{align}
where $\delta_x$ is the Dirac measure (point mass) on $x \in \R$.
It was recently shown by \cite{mailler2017measure, janson2019random} that under mild conditions, measure-valued replacement matrices such as eq. \eqref{eq:replacement_measure} have similar convergence results to traditional P\'olya urn schemes.
This allows for us to prove properties about the concentration of stake in the presence of staking derivatives, extending the analysis of \cite{fanti2019compounding}.
We further study this model by more realistic Monte Carlo simulation in \S\ref{sec:two_comp_sim}.

\subsection{Formal Properties}\label{subsec:formal}
We will prove some formal properties about the distribution of terminal staking distributions based on the update rules of equations \eqref{eq:stake_update} and \eqref{eq:replacement_measure}.
In order to prove these results we will need to make some further assumptions on the growth of the money supply and epoch lengths, which are detailed in Appendix \ref{sec:theo_assumption}.
We first note that the results of \cite[Theorem 1.4]{mailler2017measure} and \cite[Theorem 1.3]{janson2019random} guarantee\footnote{We note that \cite{bandyopadhyay2017polya} first proved results for infinite color P\'olya urns which \cite{mailler2017measure} extended to general measure-valued replacement matrices. However, both of these papers assume as `balancing' condition, akin to that of detailed balance in the MCMC literature, that effectively forces $\forall i, \;\sum_{j}\replacement_{ij} = B$ for a constant $B \in \R$. \cite{janson2019random} removes this condition, which allows for the process defined by equation \eqref{eq:replacement_measure} to be well-defined.} that under the evolution of equation \eqref{eq:stake_update}, there exists a stationary measure $\nu$ on the set of probability distributions on $\Delta^n$ such that $\bstakeprob = \lim_{h \rightarrow\infty} \stakeprob(h) \sim \nu$.
Note that all proofs of claims made can be found in Appendix \ref{sec:proofs}.

First, we make a claim about the survival probability of a validator:
\begin{claim}\label{claim:ruin_prob}
Let $\gamma$ be the probability that a validator eventually loses all of their stake. Then $\gamma = \Prob[\bstakeprobi = 0] = \frac{p_i}{1-p_i}$
\end{claim}
Note that if the slashing probability is less than 50\%, then a validator will be guaranteed to survive (e.g. have positive stake if $p_i < \frac{1}{2}$ as $\gamma$ is increasing in $p_i$ and $\gamma\vert_{p_i = \frac{1}{2}} = 1$).
Next, we consider the distribution of stake of an individual validator at block height $h$.
We study this using continuous-time embeddings of urn proceses.
These methods, pioneered in \cite[V]{athreya_ney_1972}, take a discrete time trajectory $X_i$ (such an urn process) and embed it into a continuous time process $X(t)$.
Events $\tau_1, \ldots, \tau_n, \ldots$ are drawn \emph{independently} from a memoryless distribution such that $X(\tau_i)$ represents the $i$th birth-death event.
If $X(t)$ is constructed correctly, then the laws of $\{X_i\}_{i\in\N}$ and $\{X(\tau_i)\}_{i\in\N}$ are equal in distribution.
Using a construction for an embedding from \cite{thornblad2016dominating}, we are able get an explicit distribution for $\stakedist(h)_i$ under the assumptions of this section. 
\begin{claim}\label{claim:validator_stake}
If $p_i < \frac{1}{2}$, let $\beta_i = \frac{1-p_i}{1-2p_i}$
For all $h > 0, \stakedist(h)_i = e^{(R_h - (1+\iota)p_i)h} X_i$, $ X_i\sim (1-\gamma) \Gamma(1, \frac{1}{\beta_i}) + \gamma \delta_0$ where $\Gamma(k, \theta)$ is the gamma distribution.
\end{claim}
Firstly, this claim suggests that when $p_i > \frac{1}{2}$, we should expect a validator's stake to decay exponentially towards zero.
Moreover, if all validators have the same slash probability, e.g. $\forall i,\; p_i = p$, then the expected concentration of the stake distribution is controlled by the variance of the random variable $X$ as,
    \begin{align}\label{eq:theoretical_dispersion}
    \aleph &= \Expect\left[\frac{\Vert \stakedist(h)_i\Vert_2}{\Vert \stakedist(h)_i\Vert_1}\right] = \frac{\Expect[\sum_{i=1}^n X_i^2]}{\Expect[\sum_{i=1}^n X_i]} \nonumber \\
    &= \frac{\sigma_X^2 + \mu_X^2}{\mu_X} = \sigma_X \left(\frac{\sigma_X}{\mu_X}\right) + \mu_X
    \end{align}
where $X_i$ are i.i.d. copies of $X$ and $\mu_X, \sigma^2_X$ are the mean and variance of $X$, respectively.
For this $X$ distributed via the law in Claim \ref{claim:validator_stake}, $\mu_X = \beta, \sigma_X = (1-\gamma)^2 \mu_X^2$ so $\aleph = \beta(1 + (1-\gamma)^2)$.
This effectively says that the default probability has a sizeable effect on concentration, such that when it is difficult to default ($\gamma = 0$), we expect higher concentration than when validators have a higher likelihood of ruin.
Note that the ratio of the $L^2$ to $L^1$ norm of a non-negative vector is a dissimilarity measure akin the Gini coefficient, with high concentration meaning high $\aleph$ and low concentration\footnote{If $\stakedist(h) = S_h\delta_{i,j}$, e.g. there is a dictator, then $\Expect\left[\frac{\Vert \stakedist(h)_i\Vert_2}{\Vert \stakedist(h)_i\Vert_1}\right] = 1$. If $\stakedist(h)_i = \frac{S_h}{n}$, for all $i$, then $\Expect\left[\frac{\Vert \stakedist(h)_i\Vert_2}{\Vert \stakedist(h)_i\Vert_1}\right] = \frac{1}{n}$. This ratio's similarity to the Gini coefficient can be viewed as a reasonable proxy for Gini \cite{allison1978measures}} occurring when $\aleph = \frac{1}{n}$.

We note that one can remove constant block reward assumption of assumption \ref{ass:rewards} and have inflationary rewards (in the sense of \cite{chitra2019competitive}) with an increase in complexity to the distributional equation of the claim.
Finally, we consider how the synthetic price, $\varphi_i(\stakedist(h)_i)$ behaves under certain regularity conditions:
\begin{claim}\label{claim:smale}
Suppose $\exists s \in (0, 1)$ such that $\phi$ is $L$-Lipschitz on $I = [s, 1]$ and $\exists x \in I$ such that $\phi(x) = x$. Then for all $i$, 
\[
\rho(L) = \Prob\left[ \lim_{h\rightarrow \infty} \varphi_i(\stakedist(h)_i) = x \bigg| \forall i,\, \stakedist(h)_i \in I\right] > 0
\]
with $\rho(L)$ decreasing in $L$. Further, $\exists \epsilon > 0$ such that an $\epsilon$-sized neighborhood of the fixed point $x$ will be visited infinitely often. 
\end{claim}
This claim says that if the stake distribution stays in a `safe' regime (which is defined by the Lipschitz parameter $L$), then the price of the synthetic will infinitely often visit a fixed point of $\phi$.
In the case of a staking derivative, this condition is guaranteed to hold as $\phi(1) = 1$.
However, for DeFi uses, such as the Synthetix curve in \S\ref{sec:deriv_ex}, this implies that certain synthetic prices will be visited infinitely often as trading continues.
This recurrence property for the staking derivative prices suggests that if the system avoids `unsafe' regions (e.g. regions where the synthetic price exceeds the underlying price by a bounded function of $L$), then the price process should oscillate around a fixed point of $\phi$.
If the fixed point is a peg value (e.g. \$1 for a stablecoin), then this claim suggests that choosing $\phi$ such that $L$ is small and $s$ is as close to 0 as possible, then one can bound the maximum deviations from the peg value (e.g. the fixed point $x$).

\subsection{Simulations}\label{sec:two_comp_sim}
In order to provide a more realistic understanding of how the urn model behaves, we turn to Monte Carlo simulation.
We relax Assumptions \ref{ass:epoch}, \ref{ass:deriv_borrow}, \ref{ass:selection}, and \ref{ass:rewards} and allow for agents to have a borrow probability, $\beta$, that represents their likelihood to borrow against their staked assets.
Our simulations use four ideal functionalities that do the following functions:
\begin{enumerate}
    \item $\mathsf{update\_borrowers}$: For each borrower $i$, flip a coin with probability $\beta_i$ to decide if a loan is needed. If $i$ hasn't borrowed more than $c_i \stakedist(h)_i$, borrow a random fraction of our stake that is less than the collateral limit.
    \item $\mathsf{mark\_loans\_at\_current\_height}$: Compute $\varphi_i$ using $\phi$ and equation $\eqref{eq:affine}$
    \item $\mathsf{clean\_defaulted\_loans}$: Find loans that have defaulted (e.g. $\varphi > \varphi_{\max}$) and zero the stake and the borrowing balance of the borrowers. Note that the network burns assets when this happens, reducing the money supply.
    \item $\mathsf{update\_stake\_distribution}$: Draws slashes (sampling $p_i$) and block producers (via $\stakedist$) to update the current stake distribution and increase the money supply. 
\end{enumerate}
Full algorithmic descriptions of these functionalities and parameters can be found in Appendix \ref{appendix:twocomp}.
When a validator defaults on a staking derivative --- $\stakedist(h)_i < c_i \stakedist(\beginloan)_i$ --- we set their stake to zero.
This has the effect of reducing the money supply and effectively giving all other validators an increase in future expected rewards, as $\stakeprob(h)_j$ increases for all validators $j \neq i$.
We made this choice of default policy based on those discussed within existing proposals \cite{agarwal_ojha_2019}.
These functionalities are combined into a main simulation loop, which performs Monte Carlo sampling of trajectories for $\stakedist(h)$.
The two most important variables are $\lslash$ and $\lborrow$, which represent the average probability for a validator to get slashed and to borrow via a staking derivative, respectively.

\begin{figure}
    \centering
    \includegraphics[scale=0.5]{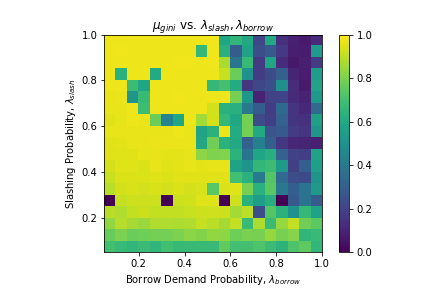}
    \includegraphics[scale=0.5]{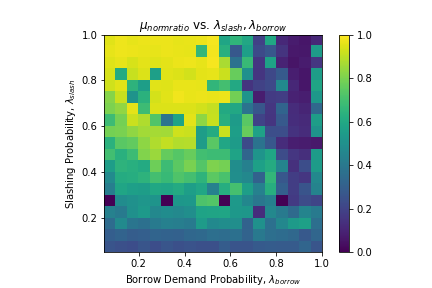}
    \caption{(Left) A heatmap of $f(a, b) = \Expect_h[\mathsf{Gini}(\stakedist(h)) | \lborrow=a, \lslash=b]$. (Right) A heatmap of $f(a, b) = \Expect_h\left[\frac{\Vert\stakedist(h)\Vert_2}{\Vert\stakedist(h)\Vert_1} \big\vert \lborrow=a, \lslash=b\right]$.
    The phase transition line between highly concentrated (yellow) and more uniform (blue) stake distribution is more clear in the Gini coefficient plot.}
    \label{fig:Gini_mean}
\end{figure}

\begin{figure}
    \centering
    \includegraphics[scale=0.5]{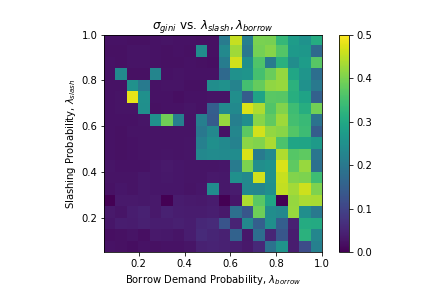}
    \includegraphics[scale=0.5]{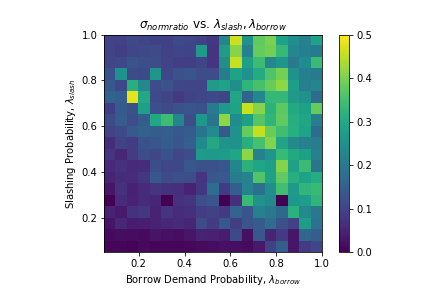}
    \caption{(Left) A heatmap of $g(a, b) = \sqrt{\Var_h[\mathsf{Gini}(\stakedist(h)) | \lborrow=a, \lslash=b]}$. (Right) A heatmap of $g(a, b) = \sqrt{\Var_h\left[\frac{\Vert\stakedist(h)\Vert_2}{\Vert\stakedist(h)\Vert_1} \big\vert \lborrow=a, \lslash=b\right]}$.
    We can see that there is a quantitative similarity between the standard deviation for the Gini coefficient and the norm ratio.
    The main feature to observe is that there is a sharp transition from no variance to a sizeable amount of variable.}
    \label{fig:Gini_std}
\end{figure}

In Figure \ref{fig:Gini_mean}, we see a heatmap of the expected Gini coefficient as a function of $\lslash$ and $\lborrow$ for an inflationary monetary policy with $\lambda = 1$.
The left-hand figure of the expected Gini coefficient shows a stark transition between highly concentrated stake distributions and much more diffuse stake distributions.
This transition line, which roughly corresponds to $\lslash = \frac{5}{4}\lborrow - 1$, shows that at high borrowing demand and relatively high slashing rates, one should expect to see a more diffuse stake distribution.
An explanation for this reduction in inequality is that once borrowing demand is high, even the larger participants end up minting staking derivatives and when they are slashed, they effectively redistribute their stake to smaller validators.

Above the critical line, $\lslash = \frac{1}{2}$, we see that the two measures are similar and report similar amounts of concentration.
As we move away from the critical line, we see that the Gini coefficient continues to stay concentrated, whereas the norm ratio disperses.
This difference between Gini and the norm ratio is expected amongst exponential family distributions \cite{allison1978measures}.
However, both figures clearly illustrate that when there is a high borrowing demand and non-trivial slashing, staking derivatives can \emph{reduce} inequality substantially.

Figure \ref{fig:Gini_std} shows the standard deviation of the Gini coefficient and the norm ratio as a function of $\lborrow, \lslash$.
Note that the scale is between $[0, 0.5]$ since the maximum variance for measures that take value in $[0,1]$ is in this range. 
We again see a phase transition between no variance and positive variance around the lines $\lborrow = \frac{4}{5}(1-\lslash)$ and $\lborrow > 0.5$, suggesting that there are large qualitative shifts in the evolution of $\stakedist$ along this line.
While the norm ratio figure doesn't have as sharp of transition as the Gini coefficient, it is clear that there is still an indication of increased turnover in this metric.
This suggests that once borrowing demand is high enough and slashing likelihoods go up, we should expect less concentration and we should expect there to be sizeable variance in the level of concentration that exists.

Combined, these results suggest that if there is a non-negligible likelihood to be slashed and a fair amount of borrowing demand, then one should expect more uniform stake distribution.
This suggests that PoS protocol designers hoping for a fairer token distribution can utilize staking derivatives to achieve these goals.
Moreover, the results for the norm ratio confirm the analysis of equation \eqref{eq:theoretical_dispersion}, which says that higher slashing probabilities lead to more uniform stake distributions.

\section{Derivative Returns and Portfolio Selection}\label{sec:stake_lend_deriv}
A natural question to ask about staking derivatives regards their effect on expected staking returns.
In order to study how derivatives impact rewards, we model the returns process of the derivative.
We then use this model to study how returns are affected when validators only borrow from the protocol (e.g. all lending is handled by the PoS protocol) and when there are external lending opportunities (akin to \cite{chitra2019competitive}).
These results are compared to traditional pricing models for fixed-income derivatives with embedded options to provide some economic intuition for the financial trade-offs faced upon the introduction of derivatives. 

\subsection{Derivative Returns Process}\label{sec:portfolio_sel}

In this section we study the return dynamics of staking derivatives and their dependence on the staking returns described in \S\ref{sec:prob_space}. We denote the returns to staking for agent $i$ at block height $h$ with $r_{s}(h)_i$, such that an agent that has staked $\stakedist(h-1)_i$ in the previous epoch will begin the current epoch with $\stakedist(h)_i=(r_{s}(h)_i+1)\stakedist(h-1)_i$. For example, returns will be positive if the agent is rewarded and negative if the agent is slashed. We define $\psi(r_s(h),h)=\phi_i \left ((r_{s}(h)_i+1))\stakedist(h-1)_i\right) \\
=\phi_i(\stakedist(h)_i)$. This emphasizes that the value of the staking derivative is a function of the stake in the current epoch and the incremental staking return. We assume agents select mean-variance-optimal portfolios \cite{markowitz1952portfolio} consisting of positions in staking and derivatives. The staking returns process can be derived from the assumptions in \S\ref{sec:prob_space}. To estimate the relevant moments of the return distribution of the derivative asset, we denote the derivative returns for epoch $t$ by $r_{d}(t)_i$, and write

\begin{equation}\label{eq:rate_def}
    r_d(t)_i=\frac{\psi_i(r_{s}(t+1)_i,t+1)}{\psi_i(r_{s}(t)_i,t)}-1
\end{equation}

\noindent In order to guarantee finite moments, we work in the safe regime where the derivative is not liquidated. Given $\psi$ smooth, we can estimate the mean return to the derivative with the second-order approximation

\begin{align} \label{eq:mean_return}
    & \mu_d(t)_i \approx B_i(t) + \frac{  \sigma_{s_i}^2}{2}C_i(t)
\end{align}

\noindent where 
\begin{align*}
    B_i(t) &= \frac{\psi_i(\mu_s(t+1)_i,t+1)-\psi_i(\mu_s(t)_i,t)}{\psi_i(\mu_s(t)_i,t)} \\ 
    C_i(t) &= \frac{1}{\psi_i(\mu_s(t),t)}\frac{\partial^2 \psi_i(\mu_s(t+1),t+1)}{\partial r_s^2}
\end{align*} 
Equation \eqref{eq:mean_return} states that one can estimate the mean derivative return by a ``base-scenario" return component given by $B_i(t)$ (for example, this can be a fixed interest rate that the borrower pays the protocol for borrowing against stake) plus a correction factor proportional to volatility.
This factor is driven by $C_i(t)$, which, following fixed-income terminology, we refer to refer to as the ``factor convexity" of $\psi$. The adjustment term in \eqref{eq:mean_return} can therefore be thought of as the ``cost of convexity" \cite[Chapter~11]{fabozzi1999advances} and is proportional to the square of volatility.
For example, for a variance in staking returns of $\sigma_{s_i}^2=20\%$, each unit increase in the factor convexity results in a 10\% gain in $\mu_d$.
This term captures the impact of non-linear effects on the derivative from staking. The most common example involves liquidation as a result of slashing, which produces a nonlinear loss to derivative borrowers. In most practical applications, the cost of convexity will be positive as liquidations compound losses from slashing. The expected losses from liquidation are increasing in the volatility of staking returns, which may indicate, for example, a higher probability of slashing.

We can also think of the cost of convexity as capturing the net of effect of embedded options in the staking derivative. An instructive analogy is that of bond options in fiat finance. We analogize $\phi$ to a bond, maturing at the end of the epoch, that the borrower issues to the protocol. If the validator is slashed within the epoch, the protocol may enforce early prepayment of the outstanding principal by seizing the validator's stake. This functions similarly to a put option on the staking asset that protects the protocol from downside losses (and produces a non-linear loss for the validator). Other examples of non-linearities may involve forms of ``credit scoring" as suggested in \cite{balance_scoring} that increase costs if borrowers become riskier during the epoch. In general, when the protocol has the right to change the terms of the loan within the epoch, the value of $\phi$ will have a positive cost of convexity for the borrower. On the other hand, if the borrower has the right to close out their loan before maturity, one can think of $\phi$ as embedding a call option for the borrower (a ``callable bond" is one that the issuer has the right to redeem prior to its maturity). If unfavorable staking returns result in higher borrowing costs during the epoch, the borrower is protected as they can buy back the derivative and close their loan. This would reduce the cost of convexity of the derivative. In this section, we only allow borrowers to rebalance at the start of the epoch and assume the protocol enforces liquidations. This results in a positive cost of convexity that leads the mean return in \eqref{eq:mean_return} to exceed the base return $B_i(t)$ due to the value of the ``options" held by the protocol. While we expect network participants to directly price these options in practice, in this work we are content with approximating their impact through the cost of convexity term in \eqref{eq:mean_return} and leave explicit pricing to future work.

For the variance terms in the derivative asset, we use a second-order approximations for the covariance term $\sigma_{sd}$ and a first-order approximation for the variance, $\sigma_d$: $\sigma_{d_i}^{2} \approx\sigma_{s_i}^2D_i^2(t)\; \sigma_{sd_i} \approx \sigma_{s_i}^2D_i(t)$ where 
\[
D_i(t)=-\frac{1}{\psi_i(\mu_s(t)_i,t)}\frac{\partial \psi_i(\mu_s(t+1)_i,t+1)}{\partial r_s}
\]
which we refer to as the ``factor duration" of $\psi$ with respect to the staking returns.
It captures the sensitivity of $\psi$ to changes in $r_s$.
Intuitively, duration measures the percentage change in $\psi$ of an infinitesimal change in $r_s$.
Given the restriction that $\phi$ be declining in its arguments, $D_i(t)$ will be non-negative, meaning the staking derivative will have positive factor duration.
In the portfolio selection context, $D_i$ is best understood as a risk measure. When $D_i=0$, the derivative has no dependence on staking return and functions similarly to a risk-free asset with a deterministic growth given by $B_i(t)$.
The derivative asset will offer higher volatility than staking when $D_i>1$ and lower volatility when $D_i<1$.
Informally, this can also be thought of as a leverage effect, dampening or magnifying exposure to volatility in staking returns.
We note that the results in the following sections assume $D_i \neq 1$. 

\subsection{Staking and Derivatives}\label{sec:portfolio_nolending}

\noindent We first consider the case where agents seek to maximize their wealth in terms of a two-component portfolio of staked and derivative assets. Agents are assumed to have varying risk preferences and asset endowments and optimize their portfolio allocations based on observed mean and variance characteristics of staking and derivative assets. We assume agents select an optimal weight vector $\boldsymbol{w}_i=[w_{s_i} , w_{d_i}]^T$ that maximizes the standard quadratic utility function 
\begin{equation}\label{eq:markowitz}
f(\boldsymbol{w}_i, \boldsymbol{\mu}_i, \lambda_i, \boldsymbol{\Sigma}_i) = \boldsymbol{w}_i^{T}\boldsymbol{\mu}_i - \frac{1}{2}\lambda_i \textbf{w}_i^{T} \boldsymbol{\Sigma}_i \boldsymbol{w}_i
\end{equation}
\noindent where $\lambda_i$ is an agent-specific risk-aversion parameter and 
\begin{equation}\label{eq:twocomp}
  \boldsymbol{\mu}_i(t)=
  \begin{bmatrix}
     \mu_s(t)_i\\
     \mu_d(t)_i \\
  \end{bmatrix}\quad
  \boldsymbol{\Sigma}_i(t)=
  \begin{bmatrix}
  \sigma_{s_i}^2 & D_i(t)\sigma_{s_i}^2\\ 
  D_i(t)\sigma_{s_i}^2 & D_i^2(t)\sigma_{s_i}^2 \\
  \end{bmatrix}
\end{equation}

\noindent We consider the constrained case where agents are restricted to solutions that satisfy $\boldsymbol{w}^{T}\boldsymbol{1} = 1$. Note that this models differs from that of \cite{chitra2019competitive} where returns to PoS and on-chain lending are assumed to be independent. Here, we explicitly model covariance between the staking and derivative returns. Furthermore, this covariance term depends on the duration of $\psi$, which can be tuned by the PoS protocol. 

\begin{claim}\label{claim:two_var_bound}
The change in portfolio weights satisfies
\begin{equation} \nonumber
\Vert \wbar_i(t+1) - \wbar_i(t) \Vert_1 \leq \vert U\left(D_i(t) \right) \vert  \vert \Delta \mu_s(t) +\Delta \mu_d(t) \vert + \bigg\vert \frac{\Delta D(t)}{(D_i(t+1)-1)(D_i(t)-1)}\bigg\vert \nonumber \times \vert \mu_s(t+1) + \mu_d(t+1) + 1 \vert \label{eq:weight_bound_1}
\end{equation}

\noindent where $\Delta(x(t))=x(t+1)-x(t)$ and

\begin{equation}
    U(D_i(t))=
    \begin{cases}
      \frac{D_i(t)}{D_i(t)-1}, & \text{if}\ D_i(t) > 1 \\
      \frac{1}{D_i(t)-1}, & \text{if} \ D_i(t) < 1 \\
    \end{cases}
  \end{equation}

\end{claim}

\noindent This claim states that the total turnover in the agent's portfolio is driven by two factors: change in mean returns and change in duration, as shown in the first and second terms of \eqref{eq:weight_bound_1} respectively.
The sensitivity to changes in mean return depends on duration, as shown in the first term in \eqref{eq:weight_bound_1}.
When duration is either very large or close to zero, then the worst-case rebalancing can be bounded by the absolute change in the mean vectors.
When duration approaches one from either direction, the system will become unstable and highly sensitive to changes in the mean vector.
Intuitively, in these situations, the variance and covariance terms of the derivative will approach the staking volatility, causing the relative attractiveness of the two assets to be highly sensitive to changes in their mean returns.
The second term in \eqref{eq:weight_bound_1} captures the change in weights due to changes in duration.
When duration is stable ($\Delta D$ is small), rebalancing will be driven primarily changes in mean returns.
When $\Delta D$ is large in proportion to $D^2(t)$, the sensitivity of the derivative to changes in the staking returns will create the possibility of a large rebalancing event.
For example, $D$ may jump as the validator approaches the collateralization ratio, since a small change in staking returns may wipe out the validator's stake.
This may prompt the validator to rebalance to avoid liquidation.
Overall, when duration is either very large or close to zero and is stable then the worst-case rebalancing will be no greater than the change in the mean returns to the two assets.
These changes in mean vectors will result from changes in the PoS protocol's policy vis-a-vis the agent.
Assuming the PoS network's monetary policy is consistent, the change in mean staking return for a given quantity staked is likely to be minimal.
Note that from \eqref{eq:mean_return}, the change in the derivative is given by 
$\Delta \mu_d(t) = \Delta B_i(t)  +  \frac{\sigma_s^2}{2} \Delta C_i(t)$, which is the change in the static return plus the change in convexity. This illustrates two approaches that the protocol can take to a change in the borrower's risk level. For example, if a borrower becomes riskier, the protocol may charge a higher `interest rate' (increasing the base return $B_i(t)$) or alternatively may increase cost of convexity for the borrower, for example by increasing collateral requirements. In isolation, either action may prompt riskier validators to rebalance to safer weights.

\subsection{Staking, Lending, and Derivatives}\label{sec:sld}
We incorporate an on-chain lending into the model of the preceding section, extending the model in \cite{chitra2019competitive} to three-asset portfolio selection.
Agents select an optimal weight vector $\boldsymbol{w}_i=[w_{s_i} , w_{d_i}, w_{{\ell}_i}]^T$ that maximizes the convex objective function $f(\boldsymbol{w}_i, \boldsymbol{\mu}_i, \lambda_i, \boldsymbol{\Sigma}_i) = \boldsymbol{w}_i^{T}\boldsymbol{\mu}_i - \frac{1}{2}\lambda_i \textbf{w}_i^{T} \boldsymbol{\Sigma}_i \boldsymbol{w}_i$ where

\begin{equation}\label{eq:mu_sigma}
  \boldsymbol{\mu}_i(t)=
  \begin{bmatrix}
     \mu_s(t)_i\\
     \mu_d(t)_i \\
     \mu_{\ell}(t)_i \\
  \end{bmatrix}\quad
  \boldsymbol{\Sigma}_i(t)=
  \begin{bmatrix}
  \sigma_{s_i}^2 & D_i(t)\sigma_{s_i}^2 & 0\\ 
  D_i(t)\sigma_{s_i}^2 & D_i^2(t)\sigma_{s_i}^2 & 0 \\
  0 & 0 & \sigma_{\ell_i}^2 \\
  \end{bmatrix}
\end{equation}

\noindent where $\mu_{\ell}$, $\sigma_{\ell}$ are the mean and volatility for on-chain lending respectively.
Staking and derivative returns are assumed to be independent of lending returns ($\sigma_{s \ell}=\sigma_{d \ell}=0$).
In this case, we have the following claim

\begin{claim}\label{claim:lending_supply} The agents allocation to the lending asset is given by $w_{\ell}(t)_i =  \frac{1}{\sigma_{\ell_i}\lambda_i}\left( \frac{IR_i(t)}{D_i(t)-1} + \mu_{\ell_i}(t) \right)$
where $IR_i(t)= B_i(t) - D_i(t)\mu_s(t)_i + \frac{  \sigma_{s_i}^2}{2}C_i(t)$
\end{claim}

\noindent The term $IR_i(t)$ can roughly be viewed as approximation of the ``instantaneous return" of $\psi$ as shown in \cite[P3]{marco2001quantitative}.
Note that the instantaneous return varies subtly from the mean return in \eqref{eq:mean_return}.
The former approximates the return to the derivative over an infinitesimal period, re-scaled to the length of the epoch.
For example, if one were to simplify $r_s$ to a continuous-time process $dr_s=\mu_sdt + \sigma_sdB(t)$ where $B(t)$ is a Brownian motion, then applying It\^o's lemma and taking the expectation $\Expect\left[\frac{d\psi(r_d(t),t)}{\psi((r_d(t),t))}\right]$ will generate the instantaneous return.
$IR_i(t)$ is comprised of a base return component given by $B_i(t)$, a drift return given by $D_i(t)\mu_t(t)$, and a diffusion term given by $\frac{ \sigma_{s_i}^2}{2}C_i(t)$.
In the case where $D>1$, lending will be increasing in $\mu_d$ and decreasing in $\mu_s$. As $\lim_{D_i(t) \to \infty} w_{\ell}(t)_i = \frac{1}{\sigma_{\ell_i}\lambda_i}\left(\mu_{\ell_i}(t)-\mu_{s_i}(t) \right)$ (users avoid the derivative and lending competes only with staking).
In the case where $D<1$ and $IR_i(t)$ is positive, the staking derivative reduces on-chain lending demand.
If $D=0$, then the derivative becomes akin to risk-free instrument and $w_{\ell}(t)_i = \frac{1}{\sigma_{\ell_i}\lambda_i}\left(\mu_{\ell_i}(t)-\mu_{d_i}(t) \right)$.
Finally note that in the case where $IR_i(t)=0$, on-chain lending is unconstrained by staking and derivative returns $w_{\ell}(t)_i = \frac{\mu_{\ell_i}(t)}{\sigma_{\ell_i}\lambda_i}$.

\subsection{$\phi$-specific results}
By restricting the functional forms of $\phi$, one can better classify the `safe' and `unsafe' regimes for a staking derivative.
First, we relate the rate of growth of $\phi$ to the staking volatility $\sigma_{s_i}^2$.
We find there is a regime where the derivative's price does not affect the expected mean returns.
When in this regime, in-protocol borrowing via $\phi$ can be easily managed by the protocol and liquidations (e.g. borrower defaults) do not significantly affect the net capital staked. 

\begin{claim}\label{claim:lipschitz_3comp}
Suppose that $\exists I \subset [0,1]$, $I$ compact such that $\phi, \partial \phi, \partial^2\phi$ are $L$-Lipschitz on $I$. Furthere, suppose that $\Prob[|\Delta \mu_s(t)_i| + |\Delta \mu_{\ell}(t)_i| > \frac{L}{\epsilon}] = 1 - 2^{-O(\epsilon)}$. Then with probability $1 - 2^{-O(\epsilon)}$, the change in mean return, $\bmu_i(t+1) - \bmu_i(t)$ can be uniformly bounded by a function that doesn't depend on $\mu_d(t+1), \mu_d(t)$ when $L < \frac{2}{\sigma_{s_i}^2}$
\end{claim}
\noindent The condition $L < \frac{2}{\sigma_{s_i}^2}$ can be though of as a `liquidity' condition.
Intuitively, this corresponds to high volatility in returns increasing the likelihood of liquidations.
When liquidations are large and happen frequently, the system tends to be more equal, which means large changes to the ROI of validators whose stake is slashed (akin to \S\ref{sec:two_comp_sim}).
Claim \ref{claim:lipschitz_3comp} says that validator ROI in the presence of derivatives is also lowered when the volatilty in staking and lending returns is high.
Suppose that $\phi(s) = \frac{1}{s^k}$ and $\phi'(s) = \frac{-k}{s^{k+1}}$.
With some algebra, Claim \ref{claim:lipschitz_3comp} then implies that ROI is
unaffected\footnote{Assume we start at $s = 1$. Then the $\partial\phi$ Lipschitz condition gives, for $s' \in (0,1)$
\[
\partial \phi(s') - \partial\phi'(1) = k\left(1-\frac{1}{s'^{k+1}}\right) \leq \frac{2}{\sigma_{s_i}^2} (1-s') 
\]
Approximating a maximum $s'$ by elision of terms of order $s'^{k+2}$ and higher yields the approximation. 
} if all borrowers take out loans of size less than $s^* \approx \left(\frac{k}{k+\frac{2}{\sigma_{s_i}^2}}\right)^{\frac{1}{k+1}}$.
When $\sigma_{s_i}^2 \gg 1, s^* \rightarrow 1$, which corresponds to validators being extremely overcollateralized and capital inefficient (see App. \ref{sec:construct_phi}, Figure \ref{fig:phi}).

Note, however, that when we are outside of the region $L < \frac{2}{\sigma_{s_i}^2}$, the derivative begins to be more important to validators to hold.
The main reasons for this are:
\begin{enumerate}
    \item Validators would rather get leverage and liquidity via the derivative (which is \emph{in-protocol} rather than externally, when $\phi$ is constructed such that $L > \frac{2}{\sigma_{s_i}^2})$
    \item The derivative is more important for validator portfolios in volatile regimes (e.g. $\sigma_{s_i}^2 \gg 1$)
    \item Validators' aggregate borrowing affects their staking ROI and they can potentially improve their ROI by borrowing (e.g. going leveraged long on the PoS asset)
\end{enumerate}
The first point suggests that $\phi$ can be chosen in a way such that most lending of a PoS asset takes place via the derivative as opposed to an out-of-protocol lending mechanism.
This allows protocol designers to avoid the pitfalls of \cite{chitra2019competitive}, where out-of-protocol lending could drain the security of a PoS network.

Claim \ref{claim:lipschitz_3comp} effectively says that a choice of $\phi$ for a staking derivative effectively places a prior belief on the maximum value that $\sigma_s^2$ can achieve.
For specific $\phi$, we conjecture that there is a sharp transition as a function of the gradient of $\phi$:
\begin{conjecture}\label{claim:transition}
Suppose that we are in the two-component model (e.g. \eqref{eq:markowitz}). Let $\sigma_{s_i}^2 > 1$, $\phi(s, k) = \frac{1}{s^k} \wedge 1$. Then $w_d(t) > 0 \iff k \leq 1$.
\end{conjecture}
This conjecture shows a sharp transition: When $k > 1$, validators do not use the borrowing facilities of the staking derivative.
We next numerically validate this conjecture and expand it to the three-component model via agent-based simulation.

\subsection{Simulation}\label{sec:fin_sim}
We extend the simulations of \S\ref{sec:two_comp_sim} to handle lending and varied risk preferences.
The main addition that we have is to model the mean returns and covariances from equation \eqref{eq:mu_sigma} and we have to replace the random borrowing policy with a Markowitz updates.
Our simulations use the convex optimization package \verb=cvxpy= to solve the Markowitz problem \cite{diamond2016cvxpy}.

\begin{figure}
    \centering
    \includegraphics[scale=0.5]{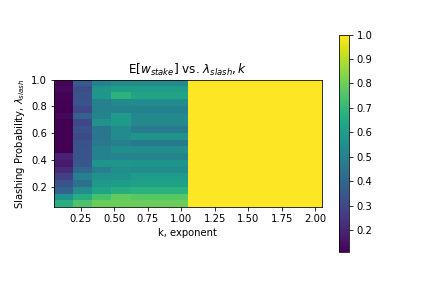}
    \includegraphics[scale=0.5]{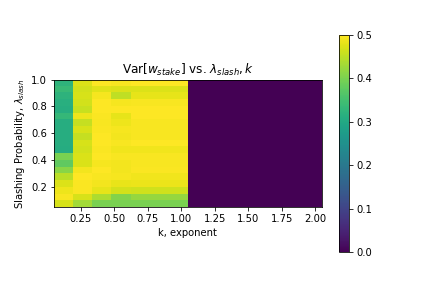}
    \caption{Plot of $\Expect_t[W_s(t)]$ (left) and $\sqrt{\Var_t[W_s(t)]}$. The phase transition suggested in Conjecture \ref{claim:transition} is clearly apparent and we can see that there is high uncertainty until we reach the low slashing and/or degree regimes.}
    \label{fig:stake_weight}
\end{figure}

\subsubsection{Two-component Model}
First, we simulate the two component model of \eqref{eq:markowitz}, eliding $r_{\ell}(t)$.
Precise details on the simulation algorithm and parameters used can be found in Appendix \ref{appendix:fin_sim}.
In Figure \ref{fig:stake_weight}, we see a heatmap of $\Expect_t[w_s(t)]$ as a function of the slashing probability $\lslash$ and the exponent $k$.
This simulation lends support to Conjecture \ref{claim:transition} and illustrates that even in high slashing regimes, borrowing becomes attractive.
Note that as $\lslash$ increases, the stake weight decays.
This is because each slash causes a liquidation (e.g. loss of the stake), leading to a decayed stake weight.
Simulations of this form can also be used to set and estimate borrowing fees, as the protocol can design fees taken upon issuing a loan to be such that $\Expect_t[w_s(t)] \geq \frac{1}{2}$.

\begin{figure}
    \centering
    \begin{subfigure}[b]{\textwidth}
     \includegraphics[width=0.45\linewidth]{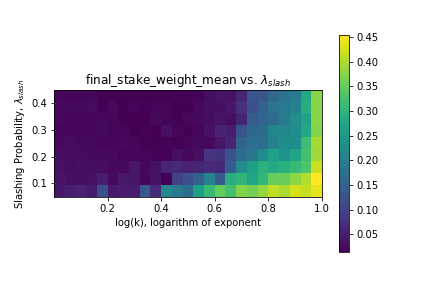}
    \includegraphics[width=0.45\linewidth]{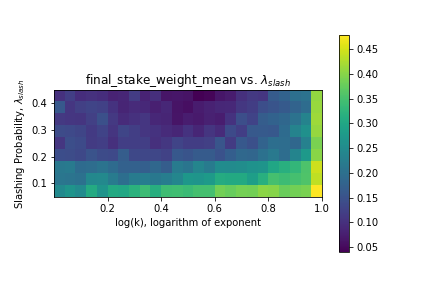}
    \includegraphics[width=0.45\linewidth]{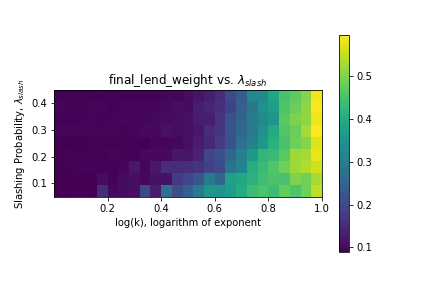}
    \includegraphics[width=0.45\linewidth]{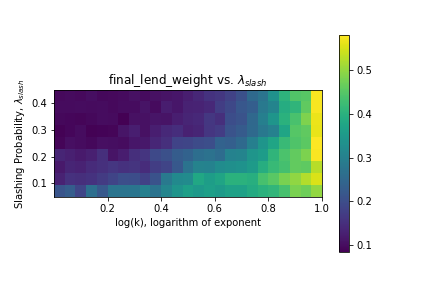}
    \caption{Portfolio weights for staked, $w_s$ (top) and lent, $w_{\ell}$ (bottom). These simulations were performed with a constant reward monetary policy and a $\iota = 0.05$ (left) and $\iota = 0.35$ (right). One can see the sharp transition, where borrowing dominates (transition happens around $\lslash \propto (\log k)^c$). Note that the higher slash fraction $\iota$, leads to more diffuse weights.}
    \label{fig:final_stake_lend_weights}
    \end{subfigure}
    
    \begin{subfigure}[b]{\textwidth}
    \includegraphics[width=0.45\linewidth]{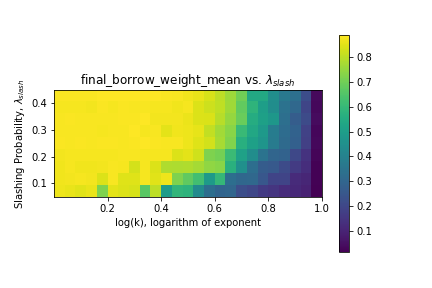}
    \includegraphics[width=0.45\linewidth]{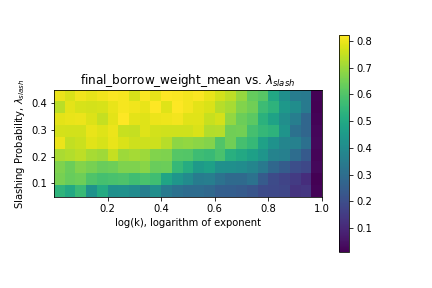}
    \includegraphics[width=0.45\linewidth]{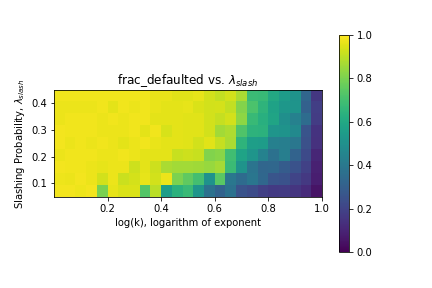}
    \includegraphics[width=0.45\linewidth]{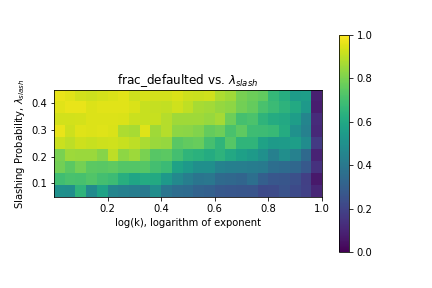}
    \caption{Portfolio weights for derivative borrowing (top) and the fraction of defaulted validators (bottom). We see the same transition from the other weights and it is clear that the higher slash fraction leads to a larger optimal region (e.g. parameter area where $w_d \approx 0.5$).}
    \label{fig:final_borrow_weight}
    \end{subfigure}
\end{figure}

\subsubsection{Three-component Model}\label{sec:three_comp_results}
\begin{figure}
    \begin{subfigure}[b]{\textwidth}
    \includegraphics[width=0.45\linewidth]{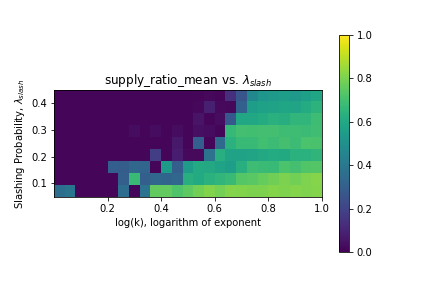}
    \includegraphics[width=0.45\linewidth]{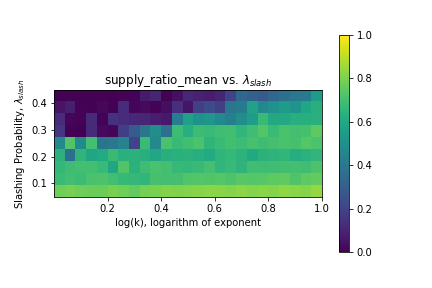}
    \caption{Supply Ratio with a deflationary monetary policy and a $\iota = 0.05$ (left) and $\iota = 0.35$ (right). We see that the higher slash fraction (right) leads to less burning of stake and a much smaller unsafe region (e.g. where the supply ratio is 0, which means that the entire money supply was burned via bad derivative lending)}
    \label{fig:supply_ratio_defl}
    \end{subfigure}
    
    \begin{subfigure}[b]{\textwidth}
    \includegraphics[width=0.45\linewidth]{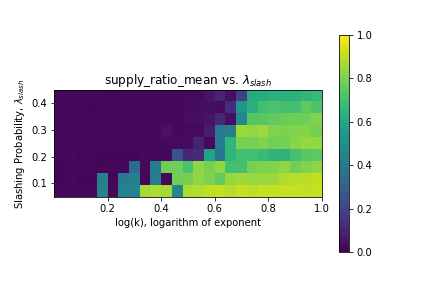}
    \includegraphics[width=0.45\linewidth]{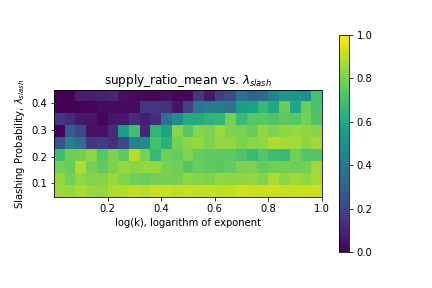}
    \caption{Supply Ratio with a constant monetary policy and a $\iota = 0.05$ (left) and $\iota = 0.35$ (right). We note the same trend in $\iota$ as the previous figure, but note that the supply ratio is higher with a non-deflationary policy. This mirrors and confirms the results of \cite{chitra2019competitive}}
    \label{fig:supply_ratio_const}
    \end{subfigure}
    
    \begin{subfigure}[b]{\textwidth}
    \includegraphics[width=0.45\linewidth]{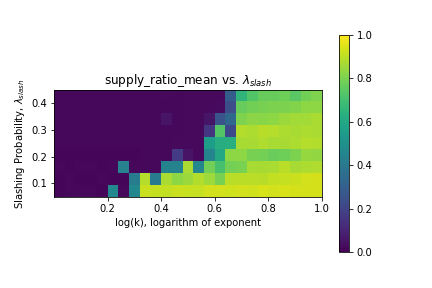}
    \includegraphics[width=0.45\linewidth]{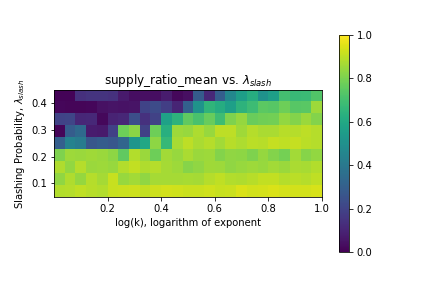}
    \caption{Supply Ratio with a inflationary monetary policy and a $\iota = 0.05$ (left) and $\iota = 0.35$ (right). Again, notice the increase in the magnitude of supply ratio, similar to \cite{chitra2019competitive}}
    \label{fig:supply_ratio_infl}
    \end{subfigure}
\end{figure}

The algorithmic description of the simulation model that includes both lending and derivatives can be found in Appendix \ref{appendix:twocomp}.
Figure \ref{fig:final_stake_lend_weights} illustrates heatmaps of the expected stake weight, $w_s(\lslash, k) = \Expect[w_s(t) | \lslash, k]$ and expected lent weights, $w_{\ell}(\lslash, k) = \Expect[w_{\ell}(t) | \lslash, k]$.
Note that the $x$-axis is logarithmic.
We can see that there is a sharp transition along a curve $\lslash = a (\log k)^c$ for some $a>  0, c > 1$ between regions of high staking and low staking.
When this threshold is crossed, we find that rational agents migrate their stake to borrowing from the staking derivative.
The borrow weight, $w_d(\lslash, k) = \Expect[w_d(t) | \lslash, k]$ can be seen in Figure \ref{fig:final_borrow_weight}.
In this figure, we see that as we flatten $\phi(s) = \frac{1}{s^k}$ by decreasing $k$, we incentivize rational validators to start borrowing against their stake in an increasingly aggressive manner.
This aggressive borrowing leads to a high number of defaults (Figure \ref{fig:final_borrow_weight}).
We quantify this further by looking at the \emph{supply ratio} $s(h)$ which is equal to $s(h) = \frac{\Vert\stakedist(h)\Vert_1}{S_h}$.
This represents ratio of the total money supply at height $h$ relative to the maximum possibly supply $S_h$.
In Figures \ref{fig:supply_ratio_defl}, \ref{fig:supply_ratio_const}, \ref{fig:supply_ratio_infl} this is illustrated for deflationary, constant, and inflationary monetary policies, respectively.
Firstly, note that when the slashing fraction $\iota$ is higher, we have a higher supply fraction for most values of $\lslash$ and $k$.
This is because the higher slash fraction $\iota$ leads to validators moving more assets to lending due to the higher default risk when using the staking derivative.
One can see this directly by noting that the lending weights in Figure \ref{fig:final_stake_lend_weights} are higher when $\iota = 0.35$.
Finally, observe that we can increase the supply ratio by having an increasingly inflationary monetary policy.
Combined, these results show that while we can reduce inequality as in \S\ref{sec:general_model} by adding derivatives, we also end up reducing the ROI for validators as a large fraction of the money supply must be burned to compensate for defaults in the derivative. 

\section{Conclusions and Future Work}
In this paper, we have explored how staking derivatives affect network security for both PoS and DeFi.
We first constructed a general framework for defining staking derivatives that encompass most of those seen in PoS and DeFi.
Then, we were able to analyze this model by using analytical techniques (with stricter assumptions) and via agent-based modeling.
We found that inequality in PoS systems can sometimes be \emph{mitigated} by the existence of staking derivatives.
The phase transition between the concentrated and non-concentrated regimes can be studied via measure-valued P\'olya urn processes, presenting new avenues to measure inequality under more realistic scenarios.

Subsequently, we estimated validators' return on investment and found that the return profile of a portfolio of derivatives and staked assets resembles a portfolio of bonds and options on bonds.
In this analysis, we found that when the derivative pricing curve was smooth and in a `safe' region far away from liquidation, we could compute the expected returns and the convexity correction.
This implies that there is an embedded option that the PoS protocol holds when derivatives are issued.
This embedded option provides many avenues for protocol developers to shape their network.
For instance, it can be tuned (via the derivative pricing function) to collect fees on derivatives, reward `good' validators, and as a form of insurance against capital flight.
In particular, our results show that there are scenarios under which the derivatives market (which is in-protocol) can become the \emph{primary} borrowing market for a staked asset.
This provides a mechanistic way for a protocol designer to avoid the capital flight of \cite{chitra2019competitive}, provided that they can choose a viable derivative pricing function.

The work presented here opens up further investigations in a number of directions.
First, the observed phase transition between the `concentrated' and `diffuse' stake distributions can likely be formally characterized.
The tools of measure-valued P\'olya processes are likely the key to proving these types of transitions and we suspect they will be of Galton-Watson type \cite[III]{athreya_ney_1972}.
Secondly, we did not probe the estimated returns in the `unsafe' regime (e.g. when expansions such as equation \ref{eq:mean_return} do not hold), which is likely where one expects to see more dramatic fluctuation in returns of the derivative.
Thirdly, we do not explore how validators and market participants explicitly price credit risk in the staking derivative and are instead satisfied with approximations for mean and variance that suffice to describe portfolio dynamics. A barrier to using the traditional credit models of \cite{merton_credit,jarrow_credit} is that the probability of default and the value that the PoS protocol can recover from the validator following default are highly dependent in the case of PoS derivatives. It is likely that the model we develop in \S\ref{sec:stake_derive} can accommodate recently-developed credit risk models \cite{urns_credit,urns_recovery} that employ urns to explicitly account for dependence between default and recovery processes \cite{urns_lgd}.
Finally, specializing this work to specific forms of the derivative pricing function, especially DeFi pricing functions (e.g. Synthetix) will likely yield new results about the unsafe region.
Analyzing empirical attacks and failures of these systems can also help shed light on how to construct optimal derivative pricing functions.

\section{Acknowledgements}
We would like to thank Yi Sun, Fabian Trottner, Matteo Liebowitz, Mario Laul, Haseeb Qureshi, Hasu, Leo Zhang, Guillermo Angeris, Michael Jordan, and Hsien-Tang Kao for helpful comments and feedback.

\bibliography{references}
\bibliographystyle{IEEEtran}

\clearpage
\onecolumn
\appendix
\section{Notation}\label{sec:notation}
We will use the following mathematical notation:
\begin{itemize}
    \item $\Delta^n$ is n-dimensional probability simplex, $\Delta^n = \{ (x_1, \ldots,
            x_n) \in \mathbb{R}^n : \sum_{i=1}^n x_i = 1, \forall i, x_i \geq 0\}$
    \item For any $x \in \mathbb{R}^n$, we define the $p$-norm as $\Vert x \Vert_p = \left(\sum_{i=1}^n |x_i|^p\right)^{1/p}$. 
    \item We turn any nonzero vector $x \in \mathbb{R}^n$ with $x \ge 0$ into a probability distribution by defining $\hat{x} = \frac{x}{\Vert x \Vert_1} \in \Delta^n$.
    \item We let $R_h$ be the block reward at height $h$ and the total money supply at time t is defined by $S_t = \sum_{h=0}^t R_h - B_t$, where $B_t$ is random variable describing the total burned token supply at time $t$
    \item $\stakedist(t) \in S_t \Delta^n$ is the unnormalized stake distribution, where $\stakedist(t)_i$ is the $i$th validators stake at time $t$
    \item $S_+^{n} \subset \R^{n\times n}$ is the cone of positive definite, symmetric matrices
    \item $\vee, \wedge$ are the standard join and meet of two elements of a lattice. For example if $a, b \in \R$, $a\wedge b = \max(a,b), a\vee b = \min(a,b)$
    \item We use standard Landau notation \cite{sipser2012introduction} on totally ordered sets $D$: Given functions $f : D \rightarrow \R, g : D \rightarrow \R$, we use the following asymptotic notations:
    \begin{itemize}
        \item $f \in O(g) \iff \exists C > 0, \forall d \in D,\, f(d) \leq C g(d)$
        \item $f \in \Omega(g) \iff \exists c > 0, \forall d \in D,\, f(d) \geq c g(d)$
        \item $f \in o(g) \iff \lim_{d \rightarrow \sup D} \frac{f(d)}{g(d)} = 0$
        \item $f \in \Theta(g) \iff f \in O(g)$ and $f \in \Omega(g)$
    \end{itemize}
\end{itemize}
We also note that the title of the paper is inspired by Bonneau's ``Why buy when you can rent?'' \cite{bonneau2016buy}.
This paper details attacks against PoW currencies that occur when there are liquid hash power derivatives (``renting''), which are the PoW equivalents of a staking derivative.

\section{Assumptions}\label{sec:assumptions}
\subsection{Common Assumptions}
The assumptions described in this section are those that apply throughout the paper.
The first two assumptions match those from previous work on PoS \cite{fanti2019compounding, chitra2019competitive}.
\begin{assumption}\label{ass:money_supply}
There is a deterministic money supply function $S_h$ that is the money supply at block height $h$. This supply function is known to all participants ahead of time and the height $h$ block reward, $R_h$, is such that $S_h = \sum_{h' \leq h} R_{h'}$.
\end{assumption}
\begin{assumption}\label{ass:static_validators}
There are a fixed number of validators, $n \in \N$, for all time.
\end{assumption}
The next assumption says that a validator's likelihood of being slashed is static.
This simplifying assumption ignores correlation between validator behavior, but does allow for variance in the likelihood of each validator being slashed.
Note that this is not a particularly strong assumption, as we can relax this significantly by assuming that all validators have upper bounds on their slashing probabilities.
\begin{assumption}\label{ass:slashing_prob}
    Each validator $i$ has a static (e.g. not changing in time) slashing probability $p_i$
\end{assumption}
Similarly, we assume that the collateral factor (e.g. maximum fraction one can borrow against their stake) is static.
This is a very reasonable assumption, as it is unlikely that collateral factors will be dynamic as they are static in DeFi \cite{kao2020compound}.
\begin{assumption}\label{ass:collat}
    Each validator $i$ has a static maximum collateral factor $c_i \in (0, 1)$, which means that the validator can borrow at most $c_i\%$ of their stake
\end{assumption}
There is also a global assumption on the slashing percentage:
\begin{assumption}
    There exists a static slashing percentage $\iota \in (0,1)$ that represents the percentage of a validators stake that is burned upon a slash
\end{assumption}
Next, we assume that validators are weakly rational in that they only commit resources to the network if they have a positive expected return:
\begin{assumption}\label{asn:weakrational}
    We assume that validators are \emph{weakly rational} in that they only have non-zero stake committed to the network at block height $h$ iff their expected returns for block $h$ are non-negative
\end{assumption}
We utilize epoch-based staking derivatives as staking derivatives for Cosmos \cite{agarwal_ojha_2019} and Tezos \cite{dexter} do this.
We also note that the liquidation period for Synthetix is a fixed time window \cite{snx_litepaper}, which provides a similar effect to this assumption.
\begin{assumption}
    Validators can only borrow from the network at the beginning of an epoch and they must repay their loans by the end of the same epoch
\end{assumption}
Finally, we utilize the simple PoS model of the staking and lending paper:
\begin{assumption}
    We will assume a simple PoS model akin to what was used in \cite{chitra2019competitive}, albeit slashing rates defined on a per validator basis.
\end{assumption}

\subsection{Assumptions for \S\ref{subsec:formal}}\label{sec:theo_assumption}
The following technical assumptions are needed to utilize generalized P\'olya urn results directly.
They can be relaxed, at the cost of making the replacement matrix significantly more complex.
\begin{assumption}\label{ass:epoch}
The epoch length $\eta$ is $1$ (e.g. one epoch, one block)
\end{assumption}

\begin{assumption}\label{ass:deriv_borrow}
All validators are maximizing their staking derivative borrows (e.g. $\forall h,\; \delta(h) = c_i \stakedist(h)_i$)
\end{assumption}

\begin{assumption}\label{ass:rewards}
We assume that $R_h$ is constant and for all $h, R_h > (1+\iota)p_i$
\end{assumption}

\begin{assumption}\label{ass:selection}
Validators can only be slashed when they are selected to be a block producer
\end{assumption}
Note is a somewhat realistic assumption, as a number of staking protocols have reduced their slashing penalties for being offline (which is the most common infraction) and increased their penalties for double signing and/or equivocation.

\section{Simulation Algorithms}\label{appendix:sim_algos}
\subsection{Model from \S\ref{sec:two_comp_sim}}\label{appendix:twocomp}
The parameters used in our Monte Carlo simulation are:
\begin{itemize}
    \item Initial stake distribution: $\stakedist(0) \leftarrow (\lceil \pi_1 \rceil, \ldots, \lceil \pi_n \rceil)$ where $\pi_i \sim \mathsf{Exp}(\lstake)$.
    \item Stake distribution at beginning of the loan: $\epochstake(\beginloan)$
    \item Collateral Factors: $c \in [0, 1]^n$ where $c_i \sim \mathsf{Beta}(\lcollateral, 1)$.
    \item Borrow Probability: $\beta \in [0, 1]^n$ where $\beta_i \sim \mathsf{Beta}(\lborrow, 1)$
    \item Slash Probability: $p \in [0, 1]^n$ where $p_i \sim \mathsf{Beta}(\lambda_{\text{slash}}, 1)$
    \item Outstanding Loans: $\ell \in \mathbb{R}_{\geq 0}^n$ that keeps track of the outstanding quantity of derivatives minted by a validator.
    \item Maximum Block Height: $h_{\max}\in \N$ is the maximum block height that a simulation was run
    \item Monetary policy parameter: $\lambda \in \R$ We used disinflationary and inflationary monetary policies characterized by $S_h = O(e^{\lambda h})$ for inflationary policies and $S_h = \frac{1-\lambda^h}{1-\lambda}$ for disinflationary policies.
    \item Derivative Pricing Function: $k \in \R_+$ is the degree in the function $\phi(s) = \frac{1}{s^k} \wedge 1$
\end{itemize}
Note that we chose an exponential stake distribution as a model of an unequal, concentration stake distribution with Gini coefficient equal to $\frac{1}{2}$, e.g. $\mathsf{Gini}(\stakedist(0)) = \frac{1}{2}$.
This distribution of wealth has been observed in western countries, including the US \cite{druagulescu2001evidence}.
Moreover, the simulations we ran were run until block height $h_{\max} = 200,000$ and we generated 100 trajectories for each combination of parameters $(\lstake, \lcollateral, \lborrow, \lslash)$.
The generative model uses four ideal functionalities:
\begin{enumerate}
    \item $\mathsf{update\_borrowers}$ 
    \item $\mathsf{mark\_loans\_at\_current\_height}$
    \item $\mathsf{clear\_defaulted\_loans}$
    \item $\mathsf{update\_stake\_distribution}$
\end{enumerate}
   
The ideal functionalities do the following:

\begin{algorithm}[H]
\caption{$\mathsf{update\_borrowers}(\ell, \beta, c, \stakedist)$}
\begin{algorithmic}
\FOR{$i \in \{1, \ldots, n_{validators}\}$}

\STATE $X \sim \text{Binomial}(\beta_i)$  
\IF{$X == 1 \wedge \ell_i < c_i \stakedist(h)_i$}
\STATE $\xi \sim \mathsf{Unif}([0, 1])$
\STATE \verb=borrow_amt_as_percentage_of_stake= $\sim \left(c_i - \frac{\ell_i}{\stakedist(h)_i}\right) \times \xi$  
\STATE $\ell_i \leftarrow$ \verb=borrow_amt_as_percentage_of_stake= $\times \pi_i$  
\ENDIF
\ENDFOR
\end{algorithmic}
\end{algorithm} 

\begin{algorithm}[H]
\caption{$\mathsf{mark\_loans\_at\_current\_height}(c, \stakedist, \ell, \epochstake, h)$}
\begin{algorithmic}
\FOR{$i \in \{1, \ldots, n_{validators}\}$}

\IF{$\ell_i > 0$}
\STATE{$b \leftarrow \frac{c_i}{1-c_i}$}
\STATE{$a \leftarrow \frac{1}{\epochstake(\beginloan)_i (c_i -1)}$ \hfill $\beginloan$ is the last epoch height, e.g. $\lfloor \frac{h}{\eta}\rfloor$}
\STATE{$\varphi_i \leftarrow \phi(a \stakedist(h)_i + b)$}
\ENDIF
\ENDFOR
\end{algorithmic}
\end{algorithm}

\begin{algorithm}[H]
\caption{$\mathsf{clean\_defaulted\_loans}(\varphi, \stakedist, h)$}
\begin{algorithmic}
\FOR{$i \in \{1, \ldots, n_{validators}\}$}

\IF{$\varphi_i > \varphi_{\max}$}
\STATE $\stakedist(h)_i \leftarrow 0$ \hfill Validator defaulted
\STATE $\beta_i \leftarrow 0$
\ENDIF
\ENDFOR
\end{algorithmic}
\end{algorithm} 

\begin{algorithm}[H]
\caption{$\mathsf{update\_stake\_distribution}(\stakedist, p, h, \iota, R_h)$}
\begin{algorithmic}
\STATE $s \leftarrow 0 \in \mathbb{R}_+^n$ \hfill Vector of validator slashings

\FOR{$i \in \{1, \ldots, n_{validators}\}$}
\STATE $s_i \sim \mathsf{Binomial}(p_i)$ \hfill Sample slashes from slashing probability distribution
\ENDFOR

\STATE $i \sim \stakeprob$  
\IF{$s_i == 0$}
\STATE $\stakedist(h+1)_i \leftarrow \stakedist(h)_i + R_h$ \hfill Add block reward to winning validator  
\ENDIF
\FOR{$j \in \{1, \ldots, n_{validators}\}$}
\IF{$s_j > 0$}
\STATE $\stakedist(h+1)_j \leftarrow (1-\iota) \stakedist(h)_j$
\ENDIF
\IF{$j \neq i \wedge s_j == 0$}
\STATE $\stakedist(h+1)_j \leftarrow \stakedist(h)_j$
\ENDIF
\ENDFOR

\end{algorithmic}
\end{algorithm} 

\begin{algorithm}[H]
\caption{Main Simulation Loop}
\begin{algorithmic}
\STATE{\# Initialize Variables}
\STATE{$h\leftarrow 0$}
\STATE{$n\leftarrow \text{Number of Agents}$}
\STATE{$\stakedist \leftarrow \pi \sim \mathsf{Exp}(\lambda_{\text{stake}})$ \hfill Initial token distribution $\pi$}
\STATE{$\epochstake \leftarrow \stakedist$ \hfill Stake distribution at last epoch}
\STATE{$\eta \leftarrow \text{Epoch Time}$}
\STATE{$\ell \leftarrow 0 \in \R^n$}
\STATE{$c \sim \prod_{i=1}^n \mathsf{Beta}(\lambda_{\text{collateral}}, 1)$ \hfill Sample collateral factors}
\STATE{$\beta \sim \prod_{i=1}^n \mathsf{Beta}(\lambda_{\text{borrow}}, 1)$ \hfill Sample borrow probabilities}
\STATE{$p \sim \prod_{i=1}^n \mathsf{Beta}(\lambda_{\text{slash}}, 1)$ \hfill Sample slashing probabilities}
\STATE{$\iota \leftarrow \text{Bond Size}$ \hfill Percentage of stake is slashed}
\STATE{$\varphi \leftarrow 1 \in \R^n$}
\STATE{$R_h \leftarrow \text{ Block Reward emission function}$}

\STATE
\STATE \# Run simulation until completion at time $T_{\max}$
\WHILE{$t < T_{\max}$}

\IF{$t \equiv 0 \text{ mod } \eta$}
\STATE{$\mathsf{update\_borrowers}(\ell, \beta, c, \stakedist)$}
\STATE{$\epochstake(\beginloan) \leftarrow \stakedist$ \hfill We annotate $\epochstake$ with $\beginloan$ for clarity}
\ENDIF

\STATE{$\mathsf{mark\_loans\_at\_current\_height}(c, \stakedist, \ell, \epochstake, \beginloan, h)$}
\STATE{$\mathsf{clear\_defaulted\_loans}(\varphi, \stakedist)$}
\STATE{$\mathsf{update\_stake\_distribution}(\stakedist, p, \iota, R_h)$}
\STATE{$\mathsf{update\_block\_reward}(h)$}
\STATE{h += 1}
\ENDWHILE

\end{algorithmic}
\end{algorithm}

\subsection{Model from \S\ref{sec:fin_sim}}\label{appendix:fin_sim}
We model the returns vector as follows
\begin{equation}
    \mathbf{r}(t)_i = 
    \begin{bmatrix}
    r_s(t)_i \\
    r_{\ell}(t)_i \\
    r_d(t)_i
    \end{bmatrix} = 
    \begin{bmatrix}
    \stakeprob(t)_i \\
    \gamma_t \\
    \frac{\psi_i(r_s(t)_i, t)}{\psi_i(r_s(t-1)_i, t-1)} - 1
    \end{bmatrix}
\end{equation}
where $\gamma_t$ is defined in \cite[\S3.2.1]{chitra2019competitive}.
In addition to the parameters described in \ref{sec:two_comp_sim}, we have the following additional parameters:
\begin{itemize}
    \item Risk aversion parameter, $\lambda_i \sim \chi^2(n)$: By having $\lambda_i$ (see \eqref{eq:markowitz}) as $\chi^2$, we allow for the expected number of risky agents to increase linearly in $n$.
    %\footnote{Recall that $\chi^2(n) \overset{d}{=} \sum_{i=1}^n Z_i$, where $Z_i \sim \mathsf{N}(0, 1)$, which gives us the correct variance scaling as $n$ increases, if we assume that each variance $\sigma_{s_i}^2$ is drawn independently from a distribution with bounded low-degree moments}
    \item Staking return variance, $\sigma_{s_i}^2$: We model the staking return variance via a stochastic process, akin to a `volatility of volatility' model from mathematical finance. We use i.i.d. Cox-Ingersoll-Ross processes:  
    \[
    d\sigma_{s_i}^2 (t) = (\kappa - \sigma_{s_i}^2(t)) dt + \xi \sigma_{s_i}(t) dB(t)
    \]
    \noindent where $dB(t)$ is the standard Brownian measure and $\kappa, \xi$ are drift and diffusion parameters.
    This model has been successfully used to model bond options (a close analogue of staking derivatives) in traditional finance \cite{heston1993closed}.
\end{itemize}
The new simulation resamples $\sigma_{s_i}^2(t)$ on each time step, updates the covariance matrix, and then recomputes the validators exposure using a Markowitz method \cite{markowitz1952portfolio}.
In this simulation, we make the following assumptions:
\begin{itemize}
    \item Each derivative borrower that isn't slashed within an epoch completely repays their loan by the end of the epoch
    \item Each on-chain loan is also repaid on epoch boundaries
\end{itemize}
The simulations of \S\ref{sec:stake_derive} can be constructed by setting the lending returns to zero.
Thus, without the loss of generality, we are only going to present the algorithm of \S\ref{sec:fin_sim}.
We replace the $\mathsf{update\_borrowers}$ functionality of the last section with a new functionality described below.
Furthermore, we introduce two new ideal functionalities that we describe below:
\begin{itemize}
    \item $\mathsf{update\_markowitz}$: This computes the optimal Markowitz portfolio using cvxpy \cite{diamond2016cvxpy}.
    \item $\mathsf{get\_returns\_and\_covariance}$: This computes the returns vector $\bmu$ and the covariance $\boSigma$ for agent $i$
\end{itemize}
We also add a lending distribution $\lenddist$ which represents the assets that the agent is supplying to an external lender (e.g. Compound).
Note that $\bmu \in \R^{n \times 3}$ and $\boSigma \in \R^{n \times 3 \times 3}$ can be thought of an arrays of validator returns and covariances.
The term $\gamma_t$ is the rate computed by the Compound smart contract, computed exactly as described in the appendix of \cite{chitra2019competitive}.
Finally, note that we assume access to an oracle that computes a sample path from a Cox-Ingersoll-Ross process.
We denote by $\mathsf{CIR}(\alpha, \beta, \sigma, t)$ a time $t$ sample from a CIR process with parameters $\alpha, \beta, \sigma$.

\begin{algorithm}[h]
\caption{$\mathsf{get\_returns\_and\_covariance}(\bmu_{prev}, \stakedist, \lenddist, \ell, i, h, \gamma_t)$}
\begin{algorithmic}
\STATE $\bmu(t) \leftarrow (0,0,0)$
\IF {$\Vert \stakedist \Vert_1 > 0$}
\STATE $r_s \leftarrow \frac{\stakedist(h)_i}{\Vert \stakedist(h) \Vert_1}$
\STATE $\mu(t)_0 \leftarrow r_s$
\STATE{$\delta \leftarrow \frac{-\phi'(r_s)}{\phi(r_s)}$ \hfill $\delta$ is duration}
\ELSE
\STATE $\mu(t)_0 \leftarrow 0$
\STATE $\delta \leftarrow 0$
\ENDIF

\IF {$\ell_i > 0$}
\STATE{$\mu(t)_1 \leftarrow \frac{\phi(r_s)}{\phi(\bmu_{prev}[i]_0)} - 1$ \hfill If $\exists i, \ell_i > 0$ then $\Vert \stakedist \Vert_1 > 0$}
\ELSE
\STATE $\mu(t)_1 \leftarrow 0$
\ENDIF

\STATE $\mu(t)_2 \leftarrow \gamma_t$

\STATE $\boSigma(t) \leftarrow \begin{bmatrix} 1 & \delta & 0 \\ \delta & \delta^2 & 0 \\ 0 & 0 & 0\end{bmatrix}$
\STATE $\boSigma(t)_{2,2} \sim \mathsf{CIR}(\alpha, \beta, \sigma, t)$
\STATE $k \sim \mathsf{CIR}(\alpha, \beta, \sigma, t)$
\STATE $\boSigma(t) \leftarrow k * \boSigma(t)$
\RETURN $\mu(t), \boSigma(t)$
\end{algorithmic}
\end{algorithm}

\begin{algorithm}
\caption{$\mathsf{update\_markowitz}(\bmu_{prev}, \stakedist, \lenddist, \ell, \gamma_t, \lambda)$}
\begin{algorithmic}
\STATE{$w \leftarrow 0 \in \R^{n \times 3}$ \hfill weight array for all agents} 
\FOR {$i \in \{1, \ldots, n_{validators}\}$}
\IF{$\stakedist(h)_i > 0$}
\STATE $\bmu, \boSigma \leftarrow \mathsf{get\_returns\_and\_covariance}(\bmu_{prev}, \stakedist, \lenddist, \ell, i, h, \gamma_t)$
\STATE{$w[i] \leftarrow \mathsf{MINIMIZE}(\bmu - \lambda w[i]^t\boSigma w[i])$ \hfill Use convex optimizer to min. strongly convex obj.}
\ENDIF
\ENDFOR
\RETURN $w$
\end{algorithmic}
\end{algorithm}

\begin{algorithm}[h!]
\caption{$\mathsf{update\_borrowers}(\bmu, \stakedist, \lenddist, \ell, \epochstake, \lambda)$}
\begin{algorithmic}
\STATE $\epochstake \leftarrow \stakedist$
\STATE{$\gamma_t \leftarrow \mathsf{co
mpute\_borrow\_rate}(\lenddist)$ \hfill Algorithm from [3]}
\STATE{$\ell \leftarrow 0 \in \R^n$\hfill Reset borrowers, assume anyone left repaid in full}
\STATE{$w \leftarrow \mathsf{update\_markowitz}(\bmu, \stakedist, \lenddist, \ell, \gamma_t, \lambda$}
\FOR{$i \in \{1, \ldots, n\}$}
\STATE{$\omega \leftarrow \stakedist(h)_i + \lenddist(h)_i$ \hfill wealth of $i$th agent}
\STATE{$\omega_s, \omega_d, \omega_l \leftarrow \omega \times w[i]_0, \omega \times w[i]_1, \omega \times w[i]_2$}
\IF{$\omega_s + \omega_d > \stakedist(h)_i$}
\STATE{$\delta \leftarrow (\omega_s + \omega_d - \stakedist(h)_i)$}
\STATE{$\lenddist(h)_i \leftarrow \lenddist(h)_i - \delta$}
\STATE{$\stakedist(h)_i \leftarrow \stakedist(h)_i + \delta$}
\ENDIF

\STATE{\# The branch below is for taking out new derivatives}
\IF{$\omega_s < \stakedist(h)_i \wedge \omega_d < \stakedist(h)_i$}
\STATE{$\ell_i \leftarrow \omega_d$}
\STATE{$\stakedist(h)_i \leftarrow \stakedist(h)_i - \omega_d$}
\ENDIF

\IF{$\omega_l > \lenddist(h)_i$}
\STATE{$\delta \leftarrow \omega_l - \lenddist(h)_i$}
\STATE{$\lenddist(h)_i \leftarrow \lenddist(h) + \delta$}
\STATE{$\stakedist(h)_i \leftarrow \stakedist(h)_i - \delta$}
\ENDIF

\ENDFOR
\end{algorithmic}
\end{algorithm}

\section{Proofs}\label{sec:proofs}

\subsection{Proof of Claim \ref{claim:ruin_prob}}
This follows directly from \cite[Lemma~2.1]{thornblad2016dominating}, once we map the setup of \S\ref{sec:stake_derive} to their problem.
In \cite{thornblad2016dominating}, one constructs a P\'olya urn such that if a ball of color $c$ is chosen, then in their notation, with probability $p_{lemma}$, a ball is added of the same color, and with probability $1-p_{lemma}$ a ball is added whose color is selected uniformly at random but is not equal to $c$.
The first scenario represents a validator being selected and receiving a block reward, whereas the latter scenario represents a validator being selected and being slashed.
As we have made assumption \ref{ass:selection}, this maps to our scenario with $p_{lemma} = 1-p_i$. 

\subsection{Proof of Claim \ref{claim:validator_stake}}
We modify the arguments made in \cite{thornblad2016dominating} and sketch how they apply to our scenario.
The P\'olya urn process $\stakedist(h)$ embeds into a continuous time Markov branching process $X(t)$ where $X(0) = \stakedist(0)$ \cite[v.9~Theorem 1]{athreya_ney_1972}.
This birth-death process adds $R_h + 1$ balls (the block reward plus the ball taken out of the urn) to $X(t)$ and removes one ball (the sampled ball from an urn).
Moreover, this process has arrival times $\tau_i \sim \mathsf{Exp}(1)$ \cite[III]{athreya_ney_1972}, which we will denote $\tau_1, \tau_2, \ldots, \tau_n, \ldots$.
At $\tau_1$, we can write a recurrence equation for $X$:
\[
X(t) = \mathbf{1}_{t \geq \tau_1} Y (X'(t-\tau_1) + X''(t-\tau_1)) + \mathbf{1}_{t < \tau_1}
\]
where $Y \sim \mathsf{Bern}(1-p_i)$.
The first term represents either a jump to zero (e.g. $Y = 0$) or a branching that starts two new processes at time $t-\tau_1$.
The second term represents the fact that we start the process with one ball (akin to the $S(0) = 1$ urn assumptions in \cite{fanti2019compounding}).
By \cite[III.9~Theorem 1]{athreya_ney_1972}, $\lim_{t\rightarrow\infty} X(t) e^{-\alpha t} = W$ and $X(t) e^{-\alpha t}$ is a non-negative martingale for $\alpha = \Expect[\replacement_i] = R_h (1-p_i) - \iota p_i$ if $\alpha > 0$ (which is assumption \ref{ass:rewards}).
By Doob's martingale convergence theorem, this converges to a limit that satisfies the distributional equation
\begin{equation}\label{eq:functional}
U = e^{-\alpha \tau_1}Y(U' + U')
\end{equation}
We now apply \cite[Lemma~2.1]{thornblad2016dominating} to receive the result.
The proof in that paper involves showing that
\begin{itemize}
    \item $(1-\gamma)\Gamma(1, \frac{1}{\beta}) + \gamma \delta_0$ satisfies equation \eqref{eq:functional} and has finite variance
    \item Showing that the operator $T_{\epsilon}$ that maps $X(t)$ to $X(t+\epsilon)$ is a contraction mapping
    \item Applying a modified Banach fixed point theorem (from \cite{roesler2001contraction}) yields that $(1-\gamma)\Gamma(1, \frac{1}{\beta}) + \gamma \delta_0$ is the unique distribution to satisfy equation \eqref{eq:functional}
\end{itemize}

\subsection{Proof of Claim \ref{claim:smale}}
This result follows from mapping the staking derivative setup to the stochastic approximation of functions of P\'olya urn processes of \cite{pemantle1991touchpoints, zhu2009nonlinear}.
Stochastic approximation, first invented by Robbins and Monro, is the same as stochastic gradient descent, which is commonly found in the machine learning literature.
If we have an urn process $X_n \in \Delta^d$, stochastic approximation considers an urn evolution,
\[
X_{n+1} = X_n + \epsilon_n \left( F(X_n) + \xi_i\right)
\]
where $\epsilon_n$ is a `step size' and $\xi_i$ is a random vector with finite moments and $F : \R^d \rightarrow \R^d$ is a vector field.
We require, via standard arguments \cite[Thm. 2.2.3]{zhu2009nonlinear} that $\epsilon_i = \Theta\left(\frac{1}{n}\right)$ in order for this process to have a non-trivial probability of convergence for continuous $F$.
Let $\bphi(\stakedist(h)) = (\varphi_1(\stakedist(h)_1), \ldots, \varphi_n(\stakedist(h)_n))$ and consider a stochastic approximation $X_n$ where $F = \bphi$.
Recall that a downcrossing fixed point of a function $F : \R_+ \rightarrow \R_+$ is point $p$ such that a) $F(p) = p$ and b) $\exists I \subset R_+$ such that $p\in I$ and $\forall p' < p, p' \in I$ we have $F(p') > p$.
Equation \eqref{eq:bdy_conditions} tells us that $\varphi_i(1) = 1$ and $\forall s < 1, \varphi'(s) \leq 0$, implying that $1$ is a downcrossing fixed point.
This implies that $(1, \ldots, 1) \in \R^n$ is a \emph{stable point} of $\bphi$ --- $\exists A \in \R^{n\times n}$ such that $\exists N \subset \R^n, (1,\ldots, 1) \in N$ such that $\forall x \in N, \langle A(x-f(1, \ldots, 1)), x - (1, \ldots, 1) \rangle \geq 0$.
From \cite[Thm. 2.2.8]{zhu2009nonlinear}, if $p$ is a stable point of a function $F$, then $\Prob[\lim_{n\rightarrow\infty} X_n \rightarrow p] > 0$, which proves the first part of the claim.

Let $\mathcal{A}_n(\epsilon) = \{ |X_n - (1, \ldots, 1)| > \epsilon \}$ be a Borel set in the Borel $\sigma$-algebra for an urn process.
Since $\Prob[\lim_{n\rightarrow\infty} X_n \rightarrow p] > 0$, $\exists \epsilon' > 0$ such that $\sum_{n} \Prob[\mathcal{A}_n(\epsilon')] < \infty$.
The Borel-Cantelli lemma immediately yields the second part of the claim.

\subsection{Proof of Claim \ref{claim:two_var_bound}}

We can show that agents select a portfolio given by

\[
\wbar_i(t) = 
\begin{bmatrix}
w_{s_i}(t) \\
w_{d_i}(t) \\
\gamma_1
\end{bmatrix}
=
\begin{bmatrix} \label{eq:w_bar_op}
    \lambda\sigma_{s_i}^2 & \lambda_iD_i(t)\sigma_{s_i}^2 & 1     \\
    \lambda_iD_i(t)\sigma_{s_i}^2 & \lambda_i D_i^2(t)\sigma_{s_i}^2 & 1     \\
    1  &  1 & 0      \\ 
\end{bmatrix}^{-1}
\left(
\begin{bmatrix}
\mu_{s_i}(t) \\
\mu_{d_i}(t) \\
0
\end{bmatrix}
+\begin{bmatrix}
0 \\
0 \\
1
\end{bmatrix}
\right)
= (\mathbf{A}(t))^{-1} (\boldsymbol{\mubar}_i(t) + \mathbf{e}_4)
\]

\noindent Define $\Delta_{D_i}(t) = D(t+1) - D(t)$ and note that the time evolution of $A(t)$ depends only on duration

\begin{align}
\mathbf{A}(t+1)=\mathbf{A}(t)+\begin{bmatrix} \label{eq:delta_mat}
    0 & \Delta_{D_i}(t) \lambda_i \sigma_{s_i}^2 & 0     \\
    \Delta_{D_i}(t) \lambda_i \sigma_{s_i}^2 &  \left(D_i^2(t)+2\Delta_{D_i}(t)D_i(t)\right)\lambda_i\sigma_{s_i}^2 & 0     \\
    0  &  0 & 0   \\ 
\end{bmatrix} =  \mathbf{A}(t) + \mathbf{\Delta}(t)
\end{align}

\noindent by the Sherman-Morrison formula for rank-$k$ updates \cite{sherman1950adjustment, hager1989updating}, we have

\begin{align}
     (\mathbf{A}(t) + \mathbf{\Delta}(t))^{-1}=\mathbf{A}(t)^{-1} + \mathbf{X}(t)
\end{align}

\noindent where

\begin{align}
     \mathbf{X}(t)=-(\mathbf{I}+\mathbf{A}(t)^{-1}\mathbf{\Delta}(t))^{-1}\mathbf{A}(t)^{-1}\mathbf{\Delta}(t)\mathbf{A}(t)^{-1}
\end{align}

\noindent It can be shown from the definitions of $\mathbf{\Delta}$ and $\mathbf{A}$ \eqref{eq:delta_mat} and \eqref{eq:w_bar_op} that

\begin{align} \label{eq:A_bound}
     \Vert \mathbf{A}(t)^{-1} \Vert_1 =   \max\left(\bigg\vert\frac{D_i(t)}{D_i(t)-1}\bigg\vert,\bigg\vert\frac{1}{1-D_i(t)}\bigg\vert,1\right) 
\end{align}

\noindent which is $\bigg\vert\frac{D_i(t)}{D_i(t)-1}\bigg\vert$ when $D_i(t)>1$ and $\bigg\vert\frac{1}{D_i(t)-1}\bigg\vert$ when $D_i(t)<1$. It can also be shown that 

\begin{align} \label{eq:X_bound}
    \Vert \mathbf{X}(t)^{-1} \Vert_1 =   \bigg\vert \frac{\Delta D(t)}{(D_i(t+1)-1)(D_i(t)-1)}\bigg\vert
\end{align}

\noindent we therefore have

\begin{align}
\Vert \wbar_i(t+1) - \wbar_i(t) \Vert_1 &= \Vert \left(\mathbf{A}_i(t)+\mathbf{\Delta}_i(t)\right)^{-1} (\bmu_{t+1} + \mathbf{e}_4) - \mathbf{A}^{-1} (\bmu_t + \mathbf{e}_4)\Vert_1 \\
& \leq \Vert \mathbf{A}^{-1}\Vert_1 \Vert (\bmu_{t+1} - \bmu_{t})\Vert_1 + \Vert \mathbf{X}\Vert_1 \Vert (\bmu_{t+1} +\mathbf{e}_4)) \Vert_1
\end{align}

\noindent Substituting \eqref{eq:A_bound} and \eqref{eq:X_bound} gives the desired result

\subsection{Proof of Claim \ref{claim:lending_supply}}

The optimal portfolio weights are given by

\[
\wbar_i(t) = 
\begin{bmatrix}
w_{s_i}(t) \\
w_{d_i}(t) \\
w_{{\ell}_i}(t) \\
\gamma
\end{bmatrix}
=
\begin{bmatrix} \label{eq:w_bar_op}
    \lambda\sigma_{s_i}^2 & \lambda_iD_i(t)\sigma_{s_i}^2 & 0& 1     \\
    \lambda_iD_i(t)\sigma_{s_i}^2 & \lambda_i D_i^2(t)\sigma_{s_i}^2 & 0& 1     \\
    0  &  0 & \lambda\sigma_{\ell_i}^2 & 1   \\ 
    1  &  1 & 1 & 0      \\ 
\end{bmatrix}^{-1}
\left(
\begin{bmatrix}
\mu_{s_i}(t) \\
\mu_{d_i}(t) \\
\mu_{\ell_i}(t) \\
0
\end{bmatrix}
+\begin{bmatrix}
0 \\
0 \\
0 \\
1
\end{bmatrix}
\right)
= (\mathbf{A}(t))^{-1} (\boldsymbol{\mubar}_i(t) + \mathbf{e}_4)
\]

\noindent where $\gamma$ is a Lagrange multiplier. The result follows directly by taking the corresponding entry of $\wbar_i(t)$.

\subsection{Proof of Claim \ref{claim:lipschitz_3comp}}

As $\psi_i, \partial\psi_i, \partial^2\psi_i$ are Lipschitz, the error term of the 2nd order Taylor expansion of $\phi$ at $r$ is bounded by $R_2(r') = O(L\sigma_{s_i}^2)$.
Thus for any $\epsilon > 0$, if $ \sigma^2_{s_i} < \frac{\epsilon}{L}$, then
\begin{equation}
R_2(r') = |\Expect[\psi_i(\mu_s(t+1), (t+1)] - \psi_i(\mu_s(t+1), (t+1)) + \partial^2_{r_s} \psi_i(t) \sigma_s^2 / 2 | < \epsilon
\end{equation}
Writing out $\mu_d(t)$ gives:

\begin{align*}
\mu_d(t) &= \frac{\Expect_t[\psi_i(r_{s_i}(t+1)_i,t+1)]}{\psi_i(r_{s_i}(t)_i,t)}-1 \\
&\approx  \frac{\psi_i(\mu_s(t+1)_i,t+1)+\frac{\sigma_{s_i}^2}{2}\partial_{r_{s_i}}^2 \psi_i(\mu_s(t+1),t+1)}{\psi_i(\mu_s(t)_i,t)}-1 \\ 
\end{align*}

Therefore, when $|\Delta\mu_s| + |\Delta\mu_{\ell}| < \frac{L}{\epsilon}$, we have
\begin{align}
    \vert \mu_d(t+1) - \mu_d(t) \vert &\leq\bigg\vert 
    \Delta B_i(t)  +  \frac{\sigma_s^2}{2} \Delta C_i(t)  \bigg\vert + 2\epsilon\nonumber \\
    &\leq \bigg\vert \Delta B_i(t) \bigg\vert + \bigg\vert \frac{\sigma_s^2}{2} \Delta C_i(t) \bigg\vert + 2\epsilon \nonumber \\
    &\leq \bigg\vert \Delta B_i(t) \bigg\vert + \frac{L \sigma_{s_i}^2}{2} \bigg\vert \frac{1}{\psi_i(\mu_s(t+1)_i,t+1)} - \frac{1}{\psi_i(\mu_s(t)_i,t)}\bigg\vert \Vert \bmu_{t+1} - \bmu_{t}\Vert_1 + 2\epsilon \label{eq:deriv}
\end{align}

\noindent Equation \eqref{eq:deriv} combined with Claim 1 of \cite{chitra2019competitive} yields:
\begin{align*}
    \Vert \bmu_{t+1} - \bmu_{t}\Vert_1 &= |\mu_s(t+1) - \mu_s(t)| + |\mu_{\ell}(t+1) - \mu_{\ell}(t)| + |\mu_d(t+1) - \mu_d(t)| \\ 
    &\leq C\left(\frac{\dstake(t)}{S_t} + \dlend(t) \right) + \bigg\vert \Delta B_i(t) \bigg\vert + \frac{L \sigma_{s_i}^2 }{2} \bigg\vert \frac{1}{\psi_i(\mu_s(t+1)_i,t+1)} - \frac{1}{\psi_i(\mu_s(t)_i,t)}\bigg\vert \Vert \bmu_{t+1} - \bmu_{t}\Vert_1 + 2\epsilon  \\
    &\leq C\left(\frac{\dstake(t)}{S_t} + \dlend(t) \right) + \vert\psi_i(\mu_s(t+2)_i,t+2) - \psi_i(\mu_s(t+1)_i,t+1)\vert + \frac{L \sigma_{s_i}^2}{2} \Vert \bmu_{t+1} - \bmu_{t}\Vert_1 + 2\epsilon \\
    & \leq C\left(\frac{\dstake(t)}{S_t} + \dlend(t) \right) + L|\dstake(t+1)| + \frac{L \sigma_{s_i}^2}{2} \Vert \bmu_{t+1} - \bmu_{t}\Vert_1 + 2\epsilon
\end{align*}
where the second inequality uses $\psi \geq 1$ and the last line uses the Lipschitz condition.
This implies that
\begin{equation}
    \left(1 - \frac{L \sigma_{s_i}^2}{2}\right) \Vert \bmu_{t+1} - \bmu_{t}\Vert_1  \leq C\left(\dstake(t) \left(1+\frac{1}{S_t}\right) + \dlend(t) \right) + 2\epsilon
\end{equation}
If $L < \frac{2}{\sigma_{s_i}^2}$, then this equation illustrates that the change in mean vector only depends on staking and lending returns.

\end{document}